\documentclass[prd,aps,twocolumn,amsfonts,showpacs,superscriptaddress,preprintnumbers,nofootinbib]{revtex4-2}

\usepackage[a4paper,margin=2cm]{geometry}

\usepackage{graphicx}
\usepackage{xcolor}
\usepackage{rotating}
\usepackage{bm,amsmath,amssymb}
\usepackage[utf8]{inputenc}
\usepackage{comment}
\usepackage{mathtools}
\usepackage{footmisc}

\usepackage{diagbox}
\usepackage{amsthm}
\usepackage{amsmath}
\usepackage{cancel}
\usepackage{mathrsfs}  
\usepackage{physics}
\usepackage{dcolumn}% Align table columns on decimal point
\usepackage{bm}% bold math

\usepackage[colorlinks=true, urlcolor=blue, linkcolor=blue, citecolor=blue, hyperindex=true, linktocpage=true]{hyperref}

\makeatletter
\DeclareRobustCommand{\cev}[1]{%
  \mathpalette\do@cev{#1}%
}
\newcommand{\do@cev}[2]{%
  \fix@cev{#1}{+}%
  \reflectbox{$\m@th#1\vec{\reflectbox{$\fix@cev{#1}{-}\m@th#1#2\fix@cev{#1}{+}$}}$}%
  \fix@cev{#1}{-}%
}
\newcommand{\fix@cev}[2]{%
  \ifx#1\displaystyle
    \mkern#23mu
  \else
    \ifx#1\textstyle
      \mkern#23mu
    \else
      \ifx#1\scriptstyle
        \mkern#22mu
      \else
        \mkern#22mu
      \fi
    \fi
  \fi
}

\newcommand{\xleft}{\cev{x}}
\newcommand{\xright}{\vec{x}}
\newcommand{\Xleft}{\cev{\boldsymbol{x}}}

\newcommand{\Xright}{\vec{\boldsymbol{x}}}
\newcommand{\Xnone}{\boldsymbol{x}}
\newcommand{\coes}{\boldsymbol{c}}
\newcommand{\Kleft}{\cev{K}}
\newcommand{\Kright}{\vec{K}}
\newcommand{\Tleft}{\cev{f}}
\newcommand{\gammaleft}{\cev{\gamma}}
\newcommand{\sigmaleft}{\cev{\sigma}}

\newcommand{\rholeft}{\cev{\rho}}
\newcommand{\Tright}{\vec{f}}
\newcommand{\gammaright}{\vec{\gamma}}
\newcommand{\sigmaright}{\vec{\sigma}}
\newcommand{\rhoright}{\vec{\rho}}

\newcommand{\fati}{{\boldsymbol{i}}}
\newcommand{\fatj}{{\boldsymbol{j}}}
\newcommand{\fatk}{{\boldsymbol{k}}}

\makeatother
\def\CA{\mathcal{A}}
\def\CB{\mathcal{B}}

\def\CG{\mathcal{G}}
\def\CH{\mathcal{H}}
\def\CI{\mathcal{I}}
\def\CJ{\mathcal{J}}
\def\CK{\mathcal{K}}

\def\CO{\mathcal{O}}
\def\CP{\mathcal{P}}
\def\CS{\mathcal{S}}

\begin{document}

%\preprint{}

\title{The algebraic structure of the gradient expansion \\in linearised classical hydrodynamics}

\author{Sa\v{s}o Grozdanov}
\affiliation{Higgs Centre for Theoretical Physics, University of Edinburgh, Edinburgh, EH8 9YL, Scotland, 
}
\affiliation{Faculty of Mathematics and Physics, University of Ljubljana, Jadranska ulica 19, SI-1000 Ljubljana, Slovenia
}

\author{Mile Vrbica}
\affiliation{Higgs Centre for Theoretical Physics, University of Edinburgh, Edinburgh, EH8 9YL, Scotland,
}

\begin{abstract}
In this work, we systematically treat the ambiguities that generically arise in the gradient expansion of any hydrodynamic theory. While these ambiguities do not affect the physical content of the equations, they induce two types of transformations in the space of transport coefficients. The first type is known as the `frame' transformations, and amounts to field redefinitions. The second type, which we introduce and formalise here, we term the `on-shell' transformations. This identifies equivalence classes of hydrodynamic theories that provide an equally valid low-energy description of the underlying microscopic theory. We show that in any (classical) theory of hydrodynamics (at arbitrary order in derivatives), the action of such transformations on the dispersion relations and two-point correlation functions is universal. We explicitly construct invariants which can then be matched to a microscopic theory. Among them are, expectedly, the low-momentum expansions of the hydrodynamic modes. The (unphysical) gapped modes can, however, be added or removed at will. Finally, we show that such transformations assign a nilpotent Lie group to every hydrodynamic theory, and discuss the related algebraic properties underlying classical hydrodynamics.
\end{abstract}

\maketitle
\tableofcontents

\section{Introduction}

The late-time, large-wavelength dynamics of a sufficiently hot theory with locally conserved currents is typically described by the theory of hydrodynamics \cite{Romatschke:2009im,Kovtun:2012rj,Romatschke:2017ejr}. Its applications range from the theory of gases and liquids, to the evolution of the quark-gluon plasma \cite{Gale:2013da,Heinz:2013th,Casalderrey-Solana:2011dxg,Busza:2018rrf,Muronga:2001zk} and the universe \cite{Weinberg2008-tu}. Hydrodynamics provides an accurate description of a system in the regime of Fourier frequencies and momenta (wavevectors) that are (much) lower than the characteristic microscopic scale of the theory. The necessary notion of a local thermal equilibrium, which underlies any hydrodynamic theory, is reflected in the fact that the grand canonical ensemble assigns conserved currents and charges (such as energy, momentum, baryonic or electric charge) their respective conjugate variables (such as temperature, velocity, and chemical potentials). A (classical) hydrodynamic theory is then constructed by promoting the conjugate variables into dynamical fields, and using the conservation equations as the equations of motion. To do that, one must express the conserved currents in terms of the conjugate variables---the hydrodynamic variables---in what are called the \emph{constitutive relations}. They take the form of a gradient expansion, in the sense that the constitutive relations involving higher powers of derivatives will more describe the physics more accurately. The simplest and standard formulations of this procedure involve either no derivatives (ideal or Euler hydrodynamics), or derivatives to first order (Navier-Stokes hydrodynamics). Theories beyond the first order may be explicitly constructed \cite{Baier:2007ix,Bhattacharyya:2007vjd,Grozdanov:2015kqa,Jaiswal:2013vta,Diles:2023tau,Diles:2020cjy,Gavassino:2024pgl}. For discussions of the convergence of the infinite order gradient expansion, see Refs.~\cite{Grozdanov:2019kge,Grozdanov:2019uhi,Grozdanov:2021jfw} and  \cite{Withers:2018srf,Heller:2020uuy,Heller:2020hnq}.

In the conventional sense, hydrodynamics is treated as a set of partial differential equations, together with the associated initial value problem. While overwhelmingly useful and accurate in predicting a breadth of physical phenomena, such formulations of hydrodynamics are plagued with theoretical difficulties, particularly in the context of relativity. For example, even the most elementary example of simple diffusion suffers from faster-than-light transport. Moreover, the textbook formulations of relativistic hydrodynamics are famously acausal, and are therefore inappropriate for solving an initial value problem \cite{Israel:1976tn,Muller:1967zza,Israel:1976efz,Hiscock:1983zz,Hiscock:1987zz,Hiscock:1985zz}. Physically, this is simply a consequence of the fact that hydrodynamics is valid in the regime of late times and large distances, whereas causality depends on the high-momentum, high-frequency behaviour of the theory (see Refs.~\cite{Heller:2022ejw,Hoult:2023clg,Gavassino:2023mad,Wang:2023csj,Abbasi:2025teu,Abbasi:2022rum} for recent discussions). In practice, such problems need to be regulated in some manner so that, for example, numerical hydrodynamic codes can run reliably for a sufficient number of steps without crashing. 

Beyond the conventional formulations of hydrodynamics at the level of equations of motion, in a more modern sense, hydrodynamics can be thought of as an effective (Schwinger-Keldysh) field theory \cite{Dubovsky:2011sj,Grozdanov:2013dba,Crossley:2015evo,Haehl:2015uoc,Jensen:2017kzi,Kovtun:2014hpa,Harder:2015nxa,Jain:2023obu}. As such, it provides a framework for the computation of response functions (correlators), which, in the low-energy regime, approximate the response functions of the full, underlying theory. The derivative expansion provides a systematic approximation scheme, yielding a series of increasingly accurate hydrodynamic response functions. The free parameters of the hydrodynamic theory---the transport coefficients, such as viscosity---are then computed by matching the hydrodynamic response function either to the experiment or the response function of the underlying microscopic theory.

The hydrodynamic gradient expansion associated with a specific microscopic theory is not unique. Ambiguities universally arise when treating hydrodynamics as a gradient expansion that terminates at a certain (non-zero) order in derivatives. Ultimately, every reliable, physical prediction of hydrodynamics must be invariant with respect to these ambiguities. %{From the point of view of various hydrodynamic ambiguities, a hydrodynamic effective field theory (EFT) should lead to reliable and observable low-energy predictions compatible with the late-time predictions of hydrodynamic simulations.}
To formalise these problems and place them within a rigorous framework, in this paper, we provide a robust and systematic mathematical formulation of the ambiguities that arise in any hydrodynamic theory. The first such ambiguity is known as the choice of a \emph{frame}, where the transformations between different `frames' are known as frame transformations. This amounts to the freedom of redefining hydrodynamic variables, which is a consequence of those variables not having any microscopic definition (meaning an expectation value of some  observable, be it in classical or quantum physics). For example, the Landau convention \cite{landau6} for relativistic hydrodynamics identifies the direction of fluid velocity with the direction of energy current, while the Eckart convention \cite{Eckart:1940te,Weinberg:1972kfs} identifies it with the direction of the particle current. Various frame choices have since been explored, especially with regards to stability and causality \cite{Bemfica:2019knx,Hoult:2020eho,Bemfica:2020zjp,Bemfica:2017wps,Kovtun:2019hdm,Kovtun:2022vas} (see also Refs.~\cite{Bhattacharyya:2024jxm,Bhattacharyya:2024tfj,Bhattacharyya:2023srn,Noronha:2021syv,Bea:2023rru,Glorioso:2017fpd,Bhambure:2024axa}). Another, less known ambiguity arises from the fact that the solutions of equations of motion at some order in the derivative expansion are expected to approximate the solutions at higher orders. This gives rise to a freedom that in practice amounts to ``plugging the equations of motion at lower orders into the higher-order hydrodynamic equations" (see e.g. Refs.~\cite{Grozdanov:2015kqa,Glorioso:2017fpd,Grozdanov:2019uhi} for discussions). In this work, we present a systematic treatment of this procedure, and refer to the respective transformations as the \emph{on-shell} transformations. If we consider the hydrodynamic equations of motion as a differential operator acting on the hydrodynamic variables, then the frame and on-shell transformations correspond to certain differential operators acting on it from right and left, respectively, followed by a truncation in the order of derivatives. Crucially, both sets of transformations can be understood as maps in the space of transport coefficients, relating theories that may have different time evolutions, but contain the same late-time, low-momentum physics. 

A better understanding of frame and on-shell transformations is desirable for several reasons. Firstly, only quantities that are invariant with respect to these transformations may be matched either to experiment or the microscopic theory. While the Kubo formulae make this a simple and clear matter in theories expanded to low orders, these issues become significantly more complicated at arbitrary orders. Secondly, different frame and on-shell choices give different time evolutions of initial data. It is therefore crucial to understand what quantities are reliably predicted by hydrodynamics, and what quantities are artefact of the truncated gradient expansion. Lastly, in analogy with gauge theories, the algebraic structure of the transformations may tell us non-trivial information about the full non-linear dynamics. In this work, we only fully answer the first of these questions, leaving the remaining two problems that pertain to non-linear extensions of our formalism to future works. 

This paper is structured as follows: in Section~\ref{sec:setup}, we describe the general form of the constitutive relations in linearised hydrodynamics. In Section~\ref{sec:correlators}, we describe how the gradient expansion acts at the level of response functions and hydrodynamic modes. In Section~\ref{sec:trans}, we define and describe frame and on-shell transformations, and then explain how they act on response functions in Section~\ref{sec:action}. In Section~\ref{sec:invariants}, we construct functions of transport coefficients that are invariant under the transformations. In Section~\ref{sec:algebra}, we present some algebraic properties of these transformations, and conclude that they generically form a nilpotent group. Finally, in Section~\ref{sec:diff} we show the formalism at work on an example of all-order diffusion. Several appendices are provided at the end, which show, in detail, all of the necessary mathematical details underpinning our construction.

\section{The constitutive relations}
\label{sec:setup}
We consider a theory with a set of $N$ independent conserved charges (such as energy, charge, or the components of the momentum), and their associated currents, denoted by $(\CJ_{(i)})^\mu$. The currents may be locally sourced in the action by some `external' gauge fields $\CA_{{(i)},\mu}$, collectively called $\CA$. The currents are assumed to be (locally) conserved, which is reflected in $N$ independent conservation equations
\begin{equation}
    \partial_t (\CJ_{(i)})^t+\partial_j (\CJ_{(i)})^j=\CS_{(i)}. \label{eq:conservation}
\end{equation}
Here, $\CS_{(i)}$ are potential source terms that are assumed to depend on first derivatives of the sources $\CA$. While we will be working with the language that is naturally adapted to relativistic theories, the present analysis is not restricted to relativistic systems. Consequently, $(\CJ_{(i)})^\mu$ are not assumed to necessarily be Lorentzian vectors.

In the grand canonical ensemble, all conserved charges are ascribed their respective conjugate variable $\mu_{(i)}$ (such as temperature, components of the velocity, or chemical potential). There are $N$ such variables, which we collectively denote as $\mu$. Assuming a static, translationally invariant background thermal state in the absence of sources, both the currents and the conjugate variables are constant in space and time. The paradigm of classical hydrodynamics entails promoting the conjugate variables to slowly-varying functions of space and time. We will refer to them as \emph{hydrodynamic variables}. It will be convenient to denote all the independent components of all conserved currents as $\CJ^a$, and all the respective components of external sources as $\CA_a$. Furthermore, we denote the $N$-dimensional tuple of all the independent hydrodynamic variables as $\mu_A$. The tuples $\CJ^a$ and $\mu_A$ obey transformation rules which reflect the symmetries and transformation rules of the currents $\CJ_{(i)}$ and $\mu_{(i)}$, respectively.\footnote{To illustrate this discussion with a concrete example, we consider a four-dimensional relativistic theory with a conserved energy-momentum tensor $T^{\mu\nu}$ and a conserved U$(1)$ current $J^\mu$. They are sourced by the metric $g_{\mu\nu}$ and the external gauge field $A_\mu$, respectively. The conservation equations are $\nabla_\mu T^{\mu\nu}=F^{\nu\lambda}J_{\lambda}$ and $\nabla_\mu J^\mu=0$, where $F^{\mu\nu}$ is the field strength tensor associated with $A_\mu$, and $\nabla_\mu$ the covariant derivative compatible with $g_{\mu\nu}$. In this case, we have $(\CJ_{(1)})^\mu=T^{\mu t}$, $(\CJ_{(2)})^\mu=T^{\mu x}$, $(\CJ_{(3)})^\mu=T^{\mu y}$, $(\CJ_{(4)})^\mu=T^{\mu z}$, and $(\CJ_{(5)})^\mu=J^{\mu}$. The associated sources $\CA$ are the respective components of the metric and the external gauge field. The source terms on the right-hand side of the conservation equations are $\CS_{(1)}=S^t$, $\CS_{(2)}=S^x$, $\CS_{(3)}=S^y$, $\CS_{(4)}=S^z$, and $\CS_{(5)}=-\Gamma^{\mu}_{\mu\lambda}J^\lambda$, where $S^\nu=F^{\nu}{}_\lambda J^\lambda-\Gamma^\mu_{\mu\alpha} T^{\alpha\nu}-\Gamma^\nu_{\mu\alpha} T^{\mu\alpha}$, with $\Gamma^{\mu}_{\nu\lambda}$ being the Christoffel symbols associated with the metric $g_{\mu\nu}$. As discussed above, the source terms depend only on first derivative of the sources. We can organise all components of the conserved currents into a tuple $\CJ^a=(T^{tt},T^{tx},T^{ty},\ldots,J^t,J^x,\ldots)$. Note that  $T^{\mu\nu}$ is symmetric, and that both $T^{\mu\nu}$ and $J^\mu$ transform as Lorentz tensors, which endows $\CJ^a$ with certain specific transformation rules. They must also obey the symmetries of the thermal state, which breaks the relativistic boost invariance, but enjoys translational and rotational invariance. Conventionally, the hydrodynamic variables in such a theory are $\mu_A=(T,u_x,u_y,u_z,\mu)$, where $T$ is the temperature, $u_i$ the velocity, and $\mu$ the chemical potential. $T$ and $\mu$ must transform as SO$(3)$ scalars, and $u_i$ as an SO$(3)$ vector, which then endows the tuple $\mu_A$ with a certain a set of transformation rules. Additional structures would arise, e.g., in the presence of a conformal symmetry.}

In this work, we limit ourselves to small (linearised) perturbations around the equilibrium state. We decompose them into Fourier modes
\begin{subequations}
\begin{align}
    \mu_{(i)}&\rightarrow \mu_{(i)}^0+\delta \mu_{(i)}e^{-i\omega t + i \vb{k}\cdot\vb{x}},\\
    \CA_{(i)}&\rightarrow                                                                                  \delta \CA_{(i)}e^{-i\omega t + i \vb{k}\cdot\vb{x}}.
\end{align}    
\end{subequations}
We assume spatial isotropy, which means that all real scalar functions are real functions of $i\omega$ and $k^2\equiv \vb k\cdot \vb k$. To write down a hydrodynamic theory, we write
\begin{equation}
\label{def:Jcurrent}
    \CJ^a(\omega,k) = \CJ_\text{ideal}^a+\delta\CJ_\text{diss}^a(\omega,k).
\end{equation}
The ideal part, which is independent of $\omega$ and $k$, is inherited from static thermodynamics, and is expressed in terms of the hydrodynamic variables via the local equation of state. The dissipative part, for which we demand that $\delta\CJ_\text{diss}^a(0,0)=0$, is expressed in terms of the gradient expansion through the \emph{constitutive relations}. This means summing over all the possible terms that are at least first and at most $n$-th order in derivatives of $\mu$ and $\CA$, and are compatible with all the symmetries and transformation rules of the conserved currents. Since we are working with a linearised theory, the number of linearly-independent terms does not grow indefinitely with the order of the gradient expansion $n$, as various tensors can be related to one another linearly through functions of $\omega$ and $k$ \cite{Grozdanov:2019uhi,Bu:2014ena,Bu:2014sia,Bu:2015ika}.  In general, we can write the constitutive relations as
\begin{equation}
    \delta\CJ^a_\text{diss}(\omega,k;\coes)=\sum_\fati c_\fati(\omega,k)\CJ^a_\fati (\omega,k;\delta\mu,\delta\CA), \label{eq:constitutive}
\end{equation}
for a finite set of linearly-independent $\CJ^a_\fati$. The functions $c_\fati(\omega,k)$ are polynomials in $\omega$ and $k$, the coefficients of which are referred to as the \emph{transport coefficients}. We collectively denote them by $\coes$. Note that the transport coefficients are subject to physical constraints. While there is a large body of literature on such constraints, see Refs.~\cite{Banerjee:2012iz,Bhattacharyya:2012nq,Jensen:2012jh,Bemfica:2020xym,Crossley:2015evo,Heller:2023jtd,Kovtun:2012rj,Gavassino:2023myj}, we keep the transport coefficients completely free in our discussion. Note also that a number of different non-linear tensors may project onto the same linearised tensor, or may even disappear in the linearised analysis. This means that the associated transport coefficients cannot be computed from linear response functions. Technical details regarding the constitutive relations can be found in Appendix~\ref{app:correlators}. 

To systematically track the order of the gradient expansion, it is convenient to introduce the \emph{truncation operation}, which truncates the order of derivatives (i.e., the degree in $\omega$ and $k$) of a given expression. Specifically, for any function $f(\omega,k)$, which is analytic around $\omega=k=0$, we define
\begin{equation}
\label{def:trun}
\qty[f(\omega,k)]_n\equiv\sum_{j=0}^n\left.\frac{1}{j!}\frac{d^j}{d\lambda^j}f(\lambda\omega,\lambda k)\right|_{\lambda=0}.
\end{equation}
The truncation operation thus only keeps powers of $\omega$ and $k$ which, when counted together, are lower than or equal to $n$. Some properties of the truncation operation are described in Appendix \ref{app:series}. The $n$-th order hydrodynamics is then defined by taking
\begin{equation}
    \CJ^a(\omega,k;\coes)=\qty[\CJ^a(\omega,k;\coes)]_n,
\end{equation}
i.e., by keeping only derivatives of $n$-th order or lower in the constitutive relations. The $N$ conservation laws \eqref{eq:conservation} constitute a closed set of equations for the $N$ hydrodynamic variables, and thus provide us with the hydrodynamic equations of motion.

Note that the $n$-th order derivative expansion may or may not be an accurate description of a system in the hydrodynamic regime when one takes into account quantum and statistical fluctuations. For example, in four spacetime dimensions, fluctuations (the so-called long time tails) are known to dominate over the contribution of second-order derivatives \cite{De_Schepper1974-od,Kovtun:2011np,Kovtun:2003vj}, whereas in three spacetime dimensions, fluctuations are dominant even in the first-order regime \cite{Forster:1977zz} (see also Refs.~\cite{Kovtun:2012rj,Grozdanov:2024fle}). Nevertheless, such fluctuations may be suppressed in (semi-classical) limits such as in the mean field theory or in theories with a large number of local degrees of freedom (e.g., in holographic large-$N$ or large central charge conformal theories). In such cases, the constitutive relations \eqref{eq:constitutive} provide a good description at any order of the truncation with the analyticity around $\omega = k = 0$ a central feature of the `classical' hydrodynamic theory. In fact, in a rotationally invariant theory, the expressions are analytic in $\omega$ and $k^2$, which leads to the dispersion relations being convergent Puiseux series. This was explained in  Refs.~\cite{Grozdanov:2019kge,Grozdanov:2019uhi}.

\section{Response functions}
\label{sec:correlators}
The equations of motion reduce to a matrix equation of the form
\begin{equation}
    F^{AB}(\omega,k;\coes)\delta\mu_B=Q^{Ab}(\omega,k;\coes)\delta\CA_b.
\end{equation}
Both $F$ and $Q$ are of degree $(n+1)$ in derivatives, i.e., $F=\qty[F]_{n+1}$ and $Q=\qty[Q]_{n+1}$. The equations of motion in the presence of sources are solved by $\delta\mu=F^{-1} Q \delta \CA$. This enables us to compute the hydrodynamic response functions, which describe the low-frequency, large-wavelength response of a conserved current to an external source, i.e., the gauge field. In this work, we treat the response functions variationally (see Refs.~\cite{Kovtun:2012rj,Herzog:2009xv}). This is in contrast with the canonical formalism of Ref.~\cite{kadanoff63}. The retarded response function $G^{ab}_\text{full}$ is then defined as
\begin{equation}
\delta\CJ^a({\delta\mu=F^{-1}Q\delta\CA})=G^{ab}_\text{full}(\omega,k)\delta \CA_b. \label{def:G}
\end{equation}
In classical hydrodynamics, it is always a rational function in $\omega$ and $k$ of the form
\begin{equation}
    G^{ab}_\text{full}(\omega,k)=\frac{B^{ab}_\text{full}(\omega,k)}{\det F(\omega,k)}. \label{eq:Gfull}
\end{equation}
Here, $B^{ab}_\text{full}(\omega,k)$ is a polynomial, the explicit, general form of which is given in Appendix \ref{app:correlators}. Since Eq.~\eqref{def:G} needs to be invariant with respect to gauge transformations of $\CA$, the components of $B_\text{full}^{ab}$ are related with one another through various Ward identities (see Appendix \ref{app:correlators} and Ref.~\cite{Herzog:2009xv}). Generically, we have (see Appendix \ref{app:invariants})
\begin{subequations}
\begin{align}
    \qty[B^{ab}_\text{full}(\omega,k)]_{N-1}&=0, & \qty[B^{ab}_\text{full}(\omega,k)]_{N}\neq 0,\\
    \qty[\det F(\omega,k)]_{N-1}&=0, & \qty[\det F(\omega,0)]_{N}\neq 0.
\end{align}    
\end{subequations}

When treating the theory from the perspective of an initial value problem, the response function of Eq.~\eqref{eq:Gfull} is the one that reflects the basic properties of the respective system of PDEs, such as causality and stability. However, when taking the perspective of a systematic derivative expansion, it is desirable for the response function of $n$-th order hydrodynamics to remain unaffected by the higher-order corrections one might add to the constitutive relation. In other words, we want the response function in $(n-1)$-th order hydrodynamics to be recoverable from $n$-th order hydrodynamics. To address this issue, we define the \emph{truncated} response function as
\begin{equation}
    G^{ab}(\omega,k)\equiv \frac{B^{ab}(\omega,k)}{P(\omega,k)}, \label{def:trunc-corr}
\end{equation}
where
\begin{subequations}
    \begin{align}
        B^{ab}(\omega,k)&=\qty[B^{ab}_\text{full}(\omega,k)]_{N+n},\\
        P(\omega,k)&=\qty[\det F(\omega,k)]_{N+n}.
    \end{align}
\end{subequations}
A response function defined this way is now independent of any higher-order corrections in the constitutive relations, and thus fits the scheme of a systematic gradient expansion (see Appendix~\ref{app:correlators}). Such a response function encapsulates all the physical, low-frequency and large-wavelength predictions of a hydrodynamic theory.\footnote{In many cases, the hydrodynamic response functions corresponding to certain degrees of freedom decouple due to symmetry (e.g., diffusive and the sound modes decouple due to isotropy, and momentum and charge dynamics decouples in a zero chemical potential state). Importantly, in such cases, the truncated response functions of the decoupled systems do not coincide with the truncated response function of the combined system. We have $\det F=\det F_1 \det F_2$, while $\qty[\det F_1 \det F_2]_{N_1+N_2+n}\neq\qty[\det F_1]_{N_1+n}\qty[\det F_2]_{N_2+n}$. Even though the truncated determinants do not decouple, they contain precisely the same amount of `reliable information' about the hydrodynamic dispersion relations as the truncated determinants of the decoupled channels (see also Section~\ref{sec:invariants}).
}

\section{Frame and on-shell transformations}
\label{sec:trans}
The gradient expansion \eqref{eq:constitutive} is riddled with ambiguities. One set of such ambiguities amounts to the freedom of choosing a \emph{frame}. Unlike the conserved currents (which are expectation values of quantum operators or functions on the phase space), the hydrodynamical variables $\mu$ have no underlying microscopic definition away from the (global) equilibrium. This means that we are free to perform a field redefinition, called a \emph{frame transformation}, as long as the equilibrium values remain intact. Concretely, this amounts to modifying the hydrodynamic variables with derivative terms that are consistent with the appropriate symmetries and transformation rules. In the linearised sense, we can write
\begin{equation}
    \delta\mu_A(\delta\mu')=\delta\mu'_A+\sum_\fatj {\xleft}_\fatj(\omega,k) \CK_{\fatj,A} (\omega,k;\delta\mu',\delta\CA). \label{def:frame}
\end{equation}
Here, $\delta\mu'_A$ are the redefined hydrodynamic variables (in a new frame) and $\CK_{\fatj,A}$ form a finite linearly-independent set of terms that are compatible with the transformation properties and symmetries of $\mu_A$. Again, because we are working with a linearised theory, there is only a finite number of (all) linearly-independent tensors. The Taylor coefficients of $\xleft_\fatj(\omega,k)$ around $\omega=k=0$, collectively denoted as $\Xleft$, are the \emph{frame coefficients} parametrising the frame transformation. (The reasons behind the overhead arrow will become clear below.) Expressing $\delta\mu$ in terms of $\delta\mu'$ introduces higher-derivative terms in the constitutive relations. By assumption, however, when working in a hydrodynamic theory to a finite order, we can truncate the higher-order terms and only keep the first $n$ orders of derivatives.\footnote{For a discussion of frame transformations which are not followed by truncation, see Refs.~\cite{Bhattacharyya:2024jxm,Bhattacharyya:2024tfj}.} Because we were using a complete basis of $\CJ^a_\fati$ in the constitutive relations \eqref{eq:constitutive}, we can now express the new constitutive relations in terms of the redefined fields $\delta\mu'$ and some new transport coefficients $\coes'$. In other words, a frame transformation induces a map on the space of transport coefficients through
\begin{equation}
    \label{def:framemap}
    \CJ^a(\delta\mu',\delta \CA;\coes')=\qty[\CJ^a(\delta\mu(\delta\mu'),\delta\CA;\coes)]_n.
\end{equation}
Such a transformation can therefore be thought of independently of the concept of a field redefinition, but rather as a transformation mapping between two hydrodynamic theories, i.e., $\coes\rightarrow \coes'$, both of which provide an equally valid effective description of an underlying microscopic theory. Frame transformations can be understood as an application of a differential operator to the equations of motion from the \emph{right}, followed by truncation. This justifies the arrow notation. In the linearised sense, this means that a frame transformation can be expressed as
\begin{equation}
\label{eq:rightmult}
    F(\omega,k;\coes ')=\qty[F(\omega,k;\coes) \Kleft(\omega,k;\Xleft) ]_{n+1},
\end{equation}
where
\begin{equation}
    \Kleft(\omega,k;0)=\Kleft(0,0;\Xleft)=\mathbb{I}.
\end{equation}
This notion translates to full, non-linear hydrodynamics.

Another set of ambiguities arises from different ways of organising the gradient expansion. At the $n$-th level of such an expansion, one can use the equations of motion of $(n-1)$-th order hydrodynamics and plug them back into the constitutive relations, with errors only appearing at an order greater than $n$ in derivatives. We refer to such transformations as the \emph{on-shell} transformations. Similarly to frame transformations, they induce a map on the space of transport coefficients. In order to treat such transformations systematically, we treat the divergences of conserved currents as new variables:
\begin{equation}
\label{def:xi}
    \delta\xi_{(i)}(\coes)=\partial_t (\CJ_{(i)})^t+\partial_j (\CJ_{(i)})^j-\CS_{(i)},
\end{equation}
with $\CJ_{(i)}$ as expressed in Eqs.~\eqref{def:Jcurrent} and \eqref{eq:constitutive}. Before imposing the equations of motion, $\xi_{(i)}(\coes)$ are just functions of derivatives, $\delta\mu$, $\delta\CA$, which do \emph{not} equal to zero. We can now write new constitutive relations \eqref{eq:constitutive}, where we include additional terms $\CH^a_\fatj(\omega,k;\delta\xi)$.\footnote{Note that, in the non-linear theory, $\CH_\fatj$ would be functions of $\mu$, $\CA$ and $\delta\xi$, as well as their respective derivatives. In the linearised theory, however, all the terms that depend on $\delta\mu$ and $\delta\CA$ are already present in the original constitutive relations. The only thing that remains are the terms linear in $\delta \xi$.} Such terms must be compatible with the symmetries and transformation properties of $\CJ^a$, and obey $\CH^a_\fatj(\omega,k;0)=0$. Because we were using a complete basis of $\CJ_\fatj^a$ in the constitutive relations \eqref{eq:constitutive}, $\CH^a_\fatj(\omega,k;\delta\xi)$ must be a linear combination of $\CJ^a_\fati$. Since using $\xi(\coes)$ as new variables introduces higher-order terms into the constitutive relations, we must truncate the new constitutive relations at $n$-th order. The new constitutive relations are expressed in terms of some new transport coefficients $\coes'$, which then amounts to our \emph{on-shell transformation}. Formally, it is defined through
\begin{equation}
\label{def:os}
    \CJ^a(\coes')=\qty[\CJ^a(\coes)+\sum_\fatj \xright_\fatj(\omega,k)\CH^a_\fatj(\omega,k;\xi(\coes))]_n.
\end{equation}
We refer to the Taylor coefficients of $\xright_{(i),j}$ around $\omega=k=0$ as the \emph{on-shell coefficients}, and denote them collectively by $\Xright$. On-shell transformations can be understood as an application of a differential operator to the equations of motion from the \emph{left}, followed by a truncation, which justifies the chosen arrow notation. In the linearised sense, this means that an on-shell transformation can be expressed as
\begin{equation}
\label{eq:leftmult}
    F(\omega,k;\coes')=\qty[\Kright(\omega,k;\Xright) F(\omega,k;\coes)]_{n+1},
\end{equation}
where
\begin{equation}
    \Kright(\omega,k;0)=\Kright(0,0;\Xright)=\mathbb{I}.
\end{equation}
This notion also translates to full, non-linear hydrodynamics.

\section{Action on response functions}
\label{sec:action}
Consider a response function (a retarded correlator) of conserved currents, $\CG^{ab}$, which accurately describes the behaviour of the theory well beyond the hydrodynamic regime (see example in Section \ref{sec:diff}). Assuming a thermal state and a set of unbroken, continuous symmetries, a classical hydrodynamic description may then be applicable in the small-$\omega$, small-$k$ regime. If this is indeed the case, then hydrodynamic response functions---rational functions of $\omega$ and $k$---are expected to approximate $\CG^{ab}$ with the appropriate choice of transport coefficients. The fact that the theory is assumed to have a hydrodynamic description is reflected in the fact that we can express
\begin{equation}
    \CG^{ab}(\omega,k)=\frac{\CB^{ab}(\omega,k)}{\CP(\omega,k)},
\end{equation}
where $\CP$ and $\CB^{ab}$ can be approximated with a (truncated) power series around $\omega=k=0$. If the hydrodynamic approximation holds up to $n$-th order, then the (truncated) hydrodynamic response function $G(\omega,k;\coes)$ is expected to take the form of \eqref{def:trunc-corr}, with
\begin{subequations}
\label{eq:hydro-approx}
    \begin{align}
        B^{ab}(\omega,k;\coes)&=\qty[\CB^{ab}(\omega,k)]_{N+n},\\
        P(\omega,k;\coes)&={\qty[\CP(\omega,k)]_{N+n}},
    \end{align}
\end{subequations}
for a choice of transport coefficients $\coes$. Let us now consider a function $p(\omega,k)$ which is analytic around $\omega=k=0$ and obeys $p(0,0)=1$. The exact response function $\CG^{ab}$ is clearly invariant under\footnote{Note that we could have relaxed the assumption of $p(0,0)=1$, which would amount to trivial rescalings and/or adding `false' hydrodynamic modes.} 
\begin{subequations}
    \begin{align}
       \CB^{ab}(\omega,k) &\rightarrow \CB^{ab}(\omega,k) p(\omega,k) , \\
       \CP(\omega,k) &\rightarrow \CP(\omega,k) p(\omega,k).
    \end{align}
\end{subequations}
It is therefore equally valid to write
\begin{equation}
    G^{ab}(\omega,k;\coes')=\frac{\qty[B^{ab}(\omega,k;\coes) p(\omega,k)]_{N+n}}{\qty[P(\omega,k;\coes) p(\omega,k)]_{N+n}}, \label{eq:trans-rule}
\end{equation}
for some alternate set of transport coefficients $\coes'$. Hydrodynamic response functions are therefore ambiguous up to the function $p$, which can be understood to be a polynomial of degree $n$. As we increase the order $n$, and thus improve the approximation, the effect of $p$ becomes less and less relevant, until it finally disappears in the strict $n\rightarrow\infty$ limit, where the series may be analitically extended beyond their radii of convergence \cite{Withers:2018srf,Grozdanov:2019uhi,Grozdanov:2022npo}.

In the language of hydrodynamics, the ambiguity of approximating such response functions with rational functions \emph{precisely} reflects the frame and on-shell ambiguities. Indeed, the action of both frame and on-shell transformations on the truncated response functions has exactly the form of Eq.~\eqref{eq:trans-rule}. This is shown in Appendices~\ref{app:frame} and \ref{app:os}. The polynomial $p$ is a function of frame and on-shell coefficients, expressed as (see Eqs.~\eqref{eq:rightmult} and \eqref{eq:leftmult})
\begin{equation}
    p(\omega,k; \Xleft,\Xright)=\det \Kleft(\omega,k;\Xleft) \det \Kright(\omega,k;\Xright), \label{def:p}
\end{equation}
and obeys
\begin{equation}
    p(0,0;\Xleft,\Xright)=p(\omega,k;0,0)=1. \label{eq:p-constraints}
\end{equation}
In other words, performing a transformation on the space of transport coefficients, is, at the level of truncated response function, equivalent to multiplying the correlator by a polynomial in both the numerator and denominator, and then truncating. While perhaps seemingly somewhat trivial, this discussion implies that frame and on-shell transformations act on response functions in the same manner. It also formalises the roles of different transformation as seen from the point of view of correlation functions.\footnote{Note that the action on full, untruncated response functions is not universal.} As a result, this gives rise to equivalence classes of response functions, with the equivalence relation defined by Eq.~\eqref{eq:trans-rule}. Generally, one expects $p$, as defined through Eq.~\eqref{def:p}, to have a very specific, prescribed structure, reflecting the structure of the matrices $K$. In actual examples, however, one can use frame and on-shell transformations to engineer an arbitrary polynomial $p$. This is also consistent with the discussion around Eq.~\eqref{eq:trans-rule}. We will henceforth assume that $p$ may be any polynomial compatible with Eq.~\eqref{eq:p-constraints}.

\section{Invariants and modes}
\label{sec:invariants}
The frame and on-shell ambiguities in formulating a hydrodynamic theory are reflected in the fact that a microscopic theory does not correspond to a unique $n$-th order hydrodynamic theory. Rather, it corresponds to an equivalence class of theories related by frame and on-shell transformations. There are two ways to tackle this. One may decide to use the frame and the on-shell transformations exhaustively to eliminate some transport coefficients. The remaining transport coefficients then unambiguously encode the physical content, and may be matched to either an underlying microscopic theory or the experiment. While this is usually done using various forms of Kubo relations, this becomes an increasingly formidable task at higher orders of the gradient expansion \cite{Grozdanov:2015kqa,Moore:2010bu,Moore:2012tc}. Although such a procedure is convenient when treating hydrodynamics as a set of PDEs with an associated initial value problem, it obscures the separation between physical content and artefacts of the truncated gradient expansion. For example, it may yield a response function which has undesirable properties (e.g. acausality, instability, or violation of the hermiticity condition $\omega \Im G^{aa}\geq 0$). Understanding the constraints on transport coefficients that arise from physical principles (such as unitarity, causality or the Onsager relations, see, e.g., Refs.~\cite{Banerjee:2012iz,Bhattacharyya:2012nq,Jensen:2012jh,Bemfica:2020xym,Crossley:2015evo,Heller:2023jtd,Kovtun:2012rj,Gavassino:2023myj}), may therefore be difficult. Alternatively, one can consider functions of transport coefficients that are invariant under frame and on-shell transformations. The benefit of this approach is in the fact that one does not have to resort to arbitrary frame choices. Furthermore, in this approach, the invariants are robust with respect to `overcounting' of the tensor structures in the constitutive relations, and do not depend on the details of the formulation of the theory. Finally, various physical constraints on transport coefficients must be reflected, in a robust sense, on the invariants of the hydrodynamic theory.

To define the invariants in a convenient way, let $\coes$ denote the set of all transport coefficients to all orders in the hydrodynamic expansion. Let $f_n(\coes)$ denote a function of $\coes$ that depends only on the coefficients that appear in the $n$-th order constitutive relations, and let $\coes'$ denote the coefficients after they had been subjected to (all-order) frame or on-shell transformations. If $f_n(\coes)=f_n(\coes')$, then we call $f_n(\coes)$ an \emph{invariant}. Furthermore, we define an \emph{invariant polynomial} as a polynomial in $\omega$ and $k$ with the coefficients that are also invariants. Defined in this manner, the invariants are now robust to higher-order corrections of constitutive relations and thus fit naturally into the derivative expansion scheme.

In the absence of sources, the equations of motion admit solutions that correspond to the poles of $G^{ab}_\text{full}$, i.e., when $\det F(\omega,k)=0$. This describes the so-called {\em spectral curves} in the complex $(\omega,k)$ space discussed in \cite{Grozdanov:2019kge,Grozdanov:2019uhi}. A special set of such solutions is the set of \emph{hydrodynamic} (gapless) modes (dispersion relations) $\widetilde\omega_i(k)$
\begin{equation}
    \det F(\widetilde\omega_i(k),k)=0, \quad \widetilde\omega_i(0)=0.
\end{equation}
By construction, there exactly $N$ such modes.\footnote{Modes may be degenerate and thus have the same dispersion relation (e.g., the $d-2$ momentum diffusion modes in isotropic $d$-dimensional spacetime). In the formalism presented here, they must be counted as distinct modes. Moreover, trivial modes with $\omega(k)=0$ must also be counted as distinct hydrodynamic modes.} We will assume that $\omega_i(k)$ are analytic around $k=0$ and therefore admit a convergent Taylor expansion in $k$.\footnote{As mentioned above, in isotropic theories, the low-$k$ expansion of $\omega_i(k)$ generically takes the form of a convergent Puiseux series in $k^{2/m}$ for integer $m$ \cite{Grozdanov:2019kge,Grozdanov:2019uhi}. In the common cases, $m$ is at most two. In fact, $m=2$ for sound modes, and the Puiseux expansion of each of the two (or, rather, a pair) of dispersion relations coincides with the Taylor expansion in $k$.}  

We may also consider the hydrodynamic modes $\omega_i(k)$ associated with the truncated determinant $P(\omega,k)$, defined through
\begin{equation}
    P(\omega_i(k),k)=0, \quad \omega_i(0)=0.
\end{equation}
Again, there are $N$ such modes, and they coincide with $\widetilde\omega_i(k)$ of the untruncated determinant in the small-$k$ expansion:
\begin{equation}
    \widetilde\omega_i(k)-\omega_i(k)=\CO(k^{n+2}).
\end{equation}
Importantly, there is no analogous equivalence for gapped modes. It is natural to introduce the expansion coefficients of the hydrodynamic dispersion relations as
\begin{equation}
    \omega_i(k)=\sum_{j=1}^{n+1}\varpi_{ij}(\coes) k^j+\CO(k^{n+2}), \label{def:varpi}
\end{equation}
where the expansion coefficients $\varpi_{ij}(\coes)$ are functions of the transport coefficients. The information contained in $\varpi_{ij}$ can be packed into the hydrodynamic part of $P$, which we define as
\begin{equation}
    P_\text{hy}(\coes)\equiv\prod_{i=1}^N (\omega-\omega_i(k)).
\end{equation}
The complementary gapped part is then defined through
\begin{equation}
\label{def:Pgap}
    P(\coes)=P_\text{hy}(\coes)P_\text{gap}(\coes),
\end{equation}
where we have omitted explicit $\omega$ and $k$ dependence. Elementary properties of series expansions (see Appendix \ref{app:series}) give the identity
\begin{equation}
  P(\coes)=\qty[\qty[P_\text{hy}(\coes)]_{N+n}\qty[P_\text{gap}(\coes)]_n]_{N+n}.
\end{equation}
The action of a transformation $p(\Xleft,\Xright)$ (see Eqs.~\eqref{eq:trans-rule} and \eqref{def:p}) can be shown to take the form (see Appendix \ref{app:invariants})
\begin{equation}
    P(\coes')=\qty[\qty[P_\text{hy}(\coes)]_{N+n} \qty[P_\text{gap}(\coes)p(\Xleft, \Xright)]_n]_{N+n}.
\end{equation}
This transformation leaves $
\qty[P_\text{hy}]_{N+n}$ invariant, which means that the expansion coefficients $\varpi_{ij}(\coes)$ are invariants, i.e., $\varpi_{ij}(\coes)=\varpi_{ij}(\coes')$. This allows us to introduce the first invariant polynomial $\CI(\omega,k;\coes)$ as follows:
\begin{equation}
    \CI(\omega,k;\coes)\equiv\qty[P_\text{hy}(\coes)]_{N+n}. \label{def:CI}
\end{equation}
We have $\CI(\omega,k;\coes')=\CI(\omega,k;\coes)$. This is in complete accordance with physical expectations---the low-$k$ expansion of hydrodynamic modes is a reliable prediction of hydrodynamics and thus cannot be affected by frame or on-shell transformations. Furthermore, high orders in momentum are not expected to be reliably described by low-order hydrodynamics. 

Alongside the gapless modes, there may also exist other solutions that correspond to gapped modes, which are, in a theory of hydrodynamics, not reflective of the gapped excitations of the underlying microscopic theory. Such modes can be added or removed by frame and on-shell transformations. For any gapped mode $\omega_i^\text{gap}(k)$ of the truncated response function, which solves $P(\omega_i^\text{gap}(k),k)=0$ and obeys $\omega_i^\text{gap}(0)\neq 0$, a transformation exists which removes it. The explicit form of such a transformation is
\begin{equation}
    p_i(\omega,k)=\qty[\frac{\omega_i^\text{gap}(0)}{\omega_i^\text{gap}(k)-\omega}]_{n}.
\end{equation}
By applying that consecutively, all the gapped modes may be removed by a transformation of the form
\begin{equation}
    p(\omega,k)=\qty[\frac{P_\text{gap}(0,0)}{P_\text{gap}(\omega,k)}]_{n},
\end{equation}
since
\begin{equation}
    \qty[P_\text{gap}]_n=\qty[f(k) \prod_i \qty[\omega-\omega_i^\text{gap}]_n]_n.
\end{equation}
Here, $f(k)$ arises generically from the factorisation of $P_\text{gap}$. We are then left with $P\propto \CI$ (see Refs.~\cite{Bhattacharyya:2024tfj,Bhattacharyya:2024jxm}, where such a choice is referred to as the `hydro frame'). Conversely, one can add gapped modes at will, which has been exploited in studies relating to causality of hydrodynamics (see e.g.~Ref.~\cite{Hoult:2023clg}). 

On the level of the response functions, the residues associated with unphysical gapped modes approach zero as $n$ is increased, and vanish in the strict $n\rightarrow \infty$ limit, where such modes cancel out completely. The same does not hold for the residues of the hydrodynamic modes. This gives rise to another set of invariant polynomials, which correspond to the physical intuition that the low-$k$ behaviour of the residues of hydrodynamics modes is reliably described by the hydrodynamic theory. By defining
\begin{equation}
    \mathcal{Y}^{ab}(\omega,k;\coes)\equiv\qty[G^{ab}(\coes)P_\text{hy}(\coes)]_{N+n}, \label{def:Jnvariant}
\end{equation}
one can show that $\mathcal{Y}^{ab}(\omega,k;\coes')=\mathcal{Y}^{ab}(\omega,k;\coes)$. Note that various components of $\mathcal{Y}^{ab}(\omega,k;\coes)$ are related to one another via Ward identities (see Appendix~\ref{app:invariants}).

\section{Algebraic properties and group structure}
\label{sec:algebra}
In this section, we remark on the fact that frame and on-shell transformations exhibit remarkably interesting and subtle algebraic structure. While we do not present any direct applications of this discussion, we entertain the idea that it might offer some interesting and non-trivial insight into the nature of hydrodynamics. Specifically, a non-linear generalisation of the present discussion may offer insight into what the robust, frame-invariant predictions of hydrodynamics are in a systematic sense. Furthermore, history has taught us that describing a physical system in terms of its redundancies may also convey a lot of valuable information. In that spirit, we proceed in a crude analogy to gauge theories, where studying the algebraic structure turned out to be immensely powerful. A related discussion exists in Ref.~\cite{Dore:2021xqq}. The technical details of this section are presented in Appendix~\ref{app:algebra}.

Let the action of a frame transformation $\Xleft$ in the space of transport coefficients be denoted by
\begin{equation}
    \coes' = \Tleft_{\Xleft}(\coes).
\end{equation}
Its form is generic, 
\begin{equation}
    c'_\fatj=\qty[c_\fatj+\sum_{\fati}\xleft_\fati
    \qty((\gammaleft_\fati)_{ \fatj}+\sum_{\fatk} (\sigmaleft_\fati)_{\fatj\fatk}c_{\fatk})]_{n-n_\fatj}, \label{eq:actionF}
\end{equation}
with $c_\fati(\omega,k)$ defined as in Eq.~\eqref{eq:constitutive} and $\xleft_{\fatj}(\omega,k)$ defined as in Eq.~\eqref{def:frame}. $(\gammaleft_\fati)_\fatj(\omega,k)$ are vectors that are related to the ideal part of the constitutive relations, while $(\sigmaleft_\fati)_{\fatj\fatk}(\omega,k)$ are matrices related to the dissipative part. Here, $n_\fatj$ corresponds to the order of derivatives of $\CJ^a_{\fatj}$ (see Appendix \ref{app:algebra} for details). The action of on-shell transformations in the space of transport coefficients is denoted by 
\begin{equation}
    \coes' = \Tright_{\Xright}(\coes)    
\end{equation}
and has a completely analogous form,
\begin{equation}
    c'_\fatj=\qty[c_\fatj+\sum_{\fati}\xright_\fati
    \qty((\gammaright_\fati)_{ \fatj}+\sum_{\fatk} (\sigmaright_\fati)_{\fatj\fatk}c_{\fatk})]_{n-n_\fatj}. \label{eq:actionO}
\end{equation}
Here, $(\gammaright_\fati)_\fatj(\omega,k)$ is again related to the ideal part of the constitutive relations, while $(\sigmaright_\fati)_{\fatj\fatk}(\omega,k)$ is related to the dissipative part. On-shell coefficients $\xright_{\fatj}(\omega,k)$ are defined in Eq.~\eqref{def:os}.

We now list several interesting properties of these transformations. Both frame and on-shell transformations obey the composition rule, i.e.,
\begin{equation}
    \exists~\Xnone'': \quad f_{\Xnone'} \circ f_{\Xnone}=f_{\Xnone''}
\end{equation}
and have an inverse
\begin{equation}
    \exists~\Xnone': \quad f_{\Xnone'}=(f_{\Xnone})^{-1}.
\end{equation}
Explicit forms are given in Appendix~\ref{app:algebra}. The transformations therefore form a Lie group, the coordinates of which are the Taylor coefficients of $x_{\fatj}(\omega,k)$, denoted collectively as $\Xnone$. This group has several generic properties that hold in any dissipative hydrodynamic theory described in this paper, and it is the goal of this section to describe them.

We introduce the group $H$ as a set of all the transformations on the space of transport coefficients, generated by frame and on-shell transformations. Some frame and on-shell actions may correspond to the same map on $\coes$ and must be identified accordingly.\footnote{Concretely, we may have $\vec f_{\Xleft}(\coes)=\cev f_{\Xright}(\coes)$ for all $\coes$ and some specific $\Xleft$ and $\Xright$. Such a transformation corresponds to a single element in $H$.} The orbits of $H$ can be then understood as equivalence classes of hydrodynamic theories, which equally accurately describe the low-energy behaviour of a corresponding microscopic theory. In other words, the space of invariants is given formally by $C / H$, where $C$ denotes the space of transport coefficients, so that $\coes \in C$. $H$ is \emph{non-compact}, which can be readily seen from the actions \eqref{eq:actionF} and \eqref{eq:actionO}. Frame and on-shell transformations both form subgroups of $H$, which we refer to as the frame group $\cev{H}$ and the on-shell group $\vec{H}$, respectively. The coordinates of $\cev{H}$ are simply the frame coefficients $\Xleft$, while the coordinates of $\vec{H}$ are the on-shell coefficients $\Xright$. Frame and on-shell transformations \emph{commute}, which is a consequence of the associativity of matrix multiplication (see Eqs.~\eqref{eq:rightmult} and \eqref{eq:leftmult}). At the level of transport coefficients, however, this property is non-trivial, and implies\footnote{We use notation by which $G \leq H$ means that $G$ is a subgroup of $H$ and $G\trianglelefteq H$ means that $G$ is a normal subgroup of $H$. For $G \leq H$ and $K \leq H$, the commutator is given by $\qty[G, K]=\qty{g^{-1}k^{-1}gk|g\in G, k\in K}$, so that $\qty[G,K]\leq H$. }
\begin{equation}
    [\vec H,\cev H]=\qty{e}, \label{algebra:commutator}
\end{equation}
where $\qty{e}$ is the trivial (identity) group. Note that this does not imply that $H$ is a direct product of ${\vec H}$ and $\cev H$, since there may be some frame transformations that coincide with some on-shell transformations. It does, however, imply that both $\vec H$ and $\cev H$ are normal subgroups of $H$
\begin{equation}
    \vec H \trianglelefteq H, \quad \cev H \trianglelefteq H. \label{algebra:normal}
\end{equation}
$H$ admits a natural sequence of subgroups associated with the order of the derivative expansion at which they act. Let us introduce the notion of a $m$-th order transformation as a transformation which is characterised by $\qty[\CJ(\coes')]_{m-1}=\qty[\CJ(\coes)]_{m-1}$, i.e., a transformation that leaves all orders less than $m$ in the constitutive relations intact. If we denote by $H_m$ the subgroup of $H$ that is generated by all the $m$-th order transformations, we have
\begin{equation}
\label{eq:algebra-props}
    H_1=H, \quad H_{n+1}=\qty{e}, \quad H_{m+1} \leq H_m.
\end{equation}
It can be shown that the commutator of an $m_1$-th order and an $m_2$-th order transformation is an $(m_1+m_2)$-th order transformation (see Appendix \ref{app:algebra}). This implies
\begin{equation}
    \qty[H_{m_1},H_{m_2}]\leq H_{m_1+m_2}, \label{algebra:commutator2}
\end{equation}
which allows us to conclude two important properties of $H$. First one is the fact that $H_{m}$ is abelian for $m>n/2$. $H$ therefore has a large abelian subgroup. This immediately implies that \emph{$H$ is always abelian for first-order hydrodynamics}. Indeed, any commutator of any two transformations in first order hydrodynamics is necessarily of at least second order in derivatives, and thus cannot appear in the constitutive relations. For an arbitrary order, $H$ needs not be abelian. It is, however, always \emph{nilpotent}. This follows from Eq.~\eqref{algebra:commutator2}, which ensures that the lower central series of $H$ terminates.\footnote{The lower central series $H^k$ of a group, defined by $H^0=H$ and $H^{k+1}=\qty[H,H^k]$, is said to terminate if $H^k=\qty{e}$ for some $k$. This defines a nilpotent group. In our case, we have $H^k \leq H_{k+1}$, which means that $H^{n}=\qty{e}$.} At the level of the Lie algebra, nilpotency implies a vanishing Killing form, which means that quadratic Casimir operators cannot be constructed. It is therefore difficult to explicitly construct invariants in a way that would be different from the one presented in Section~\ref{sec:invariants}. Since all nilpotent algebras are solvable, they have no simple subalgebras. In stark contrast with semi-simple Lie algebras, no general classification of solvable algebras is available. It is apparent from the above discussion that nilpotency emerges from the truncation step in either the frame or on-shell transformations. It is therefore a universal feature of the gradient expansion, and is expected to be present in full, non-linear hydrodynamics.

Frame and on-shell transformations can be formulated (not necessarily faithfully) in terms of $\Kright$ and $\Kleft$ matrices, introduced in Eqs.~\eqref{eq:rightmult} and \eqref{eq:leftmult}. Indeed, we have the composition rule
\begin{subequations}
\label{eq:matrixprod}
\begin{align}   \Kright(\omega,k;\Xright'')&=\qty[\Kright(\omega,k;\Xright')\Kright(\omega,k;\Xright)]_n,\\
\Kleft(\omega,k;\Xleft'')&=\qty[\Kleft(\omega,k;\Xleft)\Kleft(\omega,k;\Xleft')]_n,
\end{align}    
\end{subequations}
as well as the inverse
\begin{subequations}
\label{eq:matrixinverse}
\begin{align}
    \Kright(\omega,k;\Xright')&=\qty[\Kright^{-1}(\omega,k;\Xright)]_n,\\
    \Kleft(\omega,k;\Xleft')&=\qty[\Kleft^{-1}(\omega,k;\Xleft)]_n.
\end{align}    
\end{subequations} 

Finally, we consider the action of $H$ on the response functions. It is given by polynomials $p$ (see Eq.~\eqref{def:p}), with the composition corresponding to simple polynomial multiplication followed by truncation
\begin{equation}    p''(\omega,k)=\qty[p(\omega,k)p'(\omega,k)]_n.
\end{equation}
Here $p$, $p'$ and $p''$ are all polynomials with $p(0,0)=p'(0,0)=p''(0,0)=1$. This gives rise to an abelian Lie group $\widetilde H$, with its coordinates being the coefficients of $p$. Its definition in terms of frame and on-shell coefficients \eqref{def:p} provides a homomorphism from $H$ (which may be non-abelian) to $\widetilde H$ (which is abelian). In other words, the action of the transformations on linear response functions is always abelian.

\section{Example: diffusion at arbitrary order}
\label{sec:diff}
We illustrate the discussion of this paper on the example of simple diffusion to arbitrary order in derivative expansion. We consider a theory with spatial, parity and time-reversal invariance, as well as translational and time-translational symmetry. Boost symmetry is explicitly broken by the thermal state. We consider a simple conserved U$(1)$ current $J^\mu$ in $(1+1)$-dimensional spacetime, where $\rho\equiv J^t$ can be considered a particle density. Such a current can be sourced by an `external' U$(1)$ gauge field $A_\mu$, so that
\begin{equation}
    J^\mu=\frac{\delta S}{\delta A_\mu}.
\end{equation}
The conjugate hydrodynamic variable is the chemical potential $\mu$. We assume a background equilibrium state with $J_0^\mu=0$ and $\mu_0=0$. In the notation of Section \ref{sec:correlators}, we have
\begin{subequations}
\begin{align}
    \CJ^a&=\qty(J^t(\omega,k),J^x(\omega,k)),\\
    \CA_b&=\qty( A_t(\omega,k), A_x(\omega,k)),\\
    \mu_A&=\qty(\mu(\omega,k)),\label{example:eom}
\end{align}    
\end{subequations}
where $k_\mu=(-\omega,k)$. The quantities that are available for the construction of constitutive relations are the chemical potential $\mu(\omega,k)$, the external gauge field $A_\mu(\omega,k)$, and derivatives thereof. To ensure gauge invariance of the conserved current, the latter can only appear through the field strength tensor $F_{\mu\nu}=i k_\mu \delta A_\nu - ik_\nu \delta A_\mu$. Furthermore, we may use the static background metric $\eta_{\mu\nu}=\text{diag}(-1,1)$ and velocity $u^\mu=(1,0)$, which defines the thermal state. In the linearised theory, the most general form of the constitutive relations is
\begin{equation}
\label{example:constitutive}
    J^\mu=\chi u^\mu \delta\mu+\delta J^\mu_\text{diss},     
\end{equation}
where the ideal part is characterised by static susceptibility $\chi$
\begin{equation}
    \chi=\left.\frac{d\rho(\mu)}{d\mu}\right|_{\mu=0},
\end{equation}
and the dissipative part by four independent tensor structures
\begin{align}
\label{example:constitutive2}
    \delta J_\text{diss}^\mu(\delta\mu,\delta A)&=[(c_1(\omega,k)u^\mu+i c_2(\omega,k) k^\mu) \delta\mu+\\
    &+ (c_3(\omega,k) u_\nu+i c_4(\omega,k) k_\nu) F^{\mu\nu} \nonumber]_n.
\end{align}
We must demand that $c_1(0,0)=0$ in order to ensure that the dissipative part is zero in equilibrium. The orders of all the tensor structures $n_\fati$ are
\begin{equation}
    n_1=0, \quad n_2=1, \quad n_3=1, \quad n_4=2,
\end{equation}
so that $c_\fati(\omega,k)=\qty[c_\fati(\omega,k)]_{n-n_\fati}$. Furthermore, because the current is real and the state spatially isotropic, we must have
\begin{equation}
    c_i(\omega,k)^*=c_i(-\omega,k) \text{ and } c_i(\omega,k)=c_i(\omega,-k).
\end{equation}
The equations of motion (see Eq.~\eqref{example:eom})
\begin{equation}
    k_\mu J^\mu=0
\end{equation}
can be used to solve for $\delta\mu$ in terms of $\delta A_\mu$:
\begin{equation}
    \delta\mu(\delta A)=-\frac{\qty[(k \delta A_t+\omega \delta A_x)c_3(\omega,k) k ]_{n+1}}{\qty[ i(\chi+c_1(\omega,k))\omega+c_2(\omega,k)(k^2-\omega^2)]_{n+1}}.
\end{equation}
Since we have only one conservation equation, we have only one hydrodynamic mode---the diffusive mode. Therefore, $N=1$. Note that the tensor structure associated with $c_4$ is transverse with respect to $k_\mu$ and therefore does not enter into the equations of motion. Nevertheless, we may leave it in the constitutive relations as the formalism presented here is robust with respect to redundant tensor structures in the constitutive relations. The full retarded response functions, defined as
\begin{equation}
    J^\mu(\delta\mu(\delta A),\delta A)=G^{\mu\nu}\delta A_\nu,
\end{equation}
are related to one another by the Ward identities:
\begin{equation}
\label{eq:ward}
    \omega^2 G^{tt}=\omega k G^{tx}=\omega k G^{xt}=k^2 G^{xx}.
\end{equation}
It is therefore enough to consider only $G^{tt}$, which is given by
\begin{equation}
    G^{tt}=\frac{B^{tt}(\omega,k)}{P(\omega,k)},
\end{equation}
where
\begin{subequations}
\label{example:P}
\begin{align}
     B^{tt}(\omega,k)&=[-k^2 (c_3(\chi+c_1+i c_2 \omega)-c_4 P)]_{n+1},\label{example:B}\\
    P(\omega,k)&=\qty[-i(\chi+c_1)\omega+c_2(\omega^2-k^2)]_{n+1}. 
\end{align}    
\end{subequations}
We note that in this simple example, there is no difference between the full and the truncated response functions. 

We now consider frame transformations by redefining the out-of-equilibrium chemical potential. The most general redefinition has the form
\begin{equation}
\delta\mu(\delta\mu')=\qty[(1+\xleft_1(\omega,k)) \delta \mu'+i\xleft_2(\omega,k) k_\mu u_\nu F^{\mu\nu}]_n,
\end{equation}
where we demand that $\xleft_1(0,0)=0$. Plugging the definition into the constitutive relations \eqref{example:constitutive}, this induces a map in the space of transport coefficients through
\begin{equation}
    \qty[J^\mu(\mu';\coes')]_n=\qty[J^\mu(\mu(\mu'),\coes)]_n.
\end{equation}
The explicit map is
\begin{subequations}
\label{example:frame}
\begin{align}
    c_1'&=\qty[c_1 +(\chi+c_1)\xleft_1]_n,\\
    c_2'&=\qty[c_2 +c_2\xleft_1]_{n-1},\\
    c_3'&=\qty[c_3-(i(\chi+c_1)\omega-c_2(\omega^2-k^2))\xleft_2]_{n-1},\\
    c_4'&=\qty[c_4+(\chi+c_1+ic_2\omega)\xleft_2]_{n-2}.
\end{align}    
\end{subequations}
Turning to the on-shell transformations, we define (see Eq.~\eqref{def:xi})
\begin{align}
    \xi \equiv i k_\mu J^\mu=\qty[P(\omega,k)\delta\mu-c_3 k(k \delta A_t+\omega \delta A_x)]_{n+1} .
\end{align}
We can now add new tensors to the constitutive relations, using $\xi$ as an independent scalar together with $\mu$ and $A_\mu$. There are two new tensors that can be added, namely,
\begin{equation}
    J^\mu(\coes')=\qty[J^\mu(\coes)+(\xright_1(\omega,k)u^\mu + i \xright_2(\omega,k) k^\mu)\xi]_{n}.
\end{equation}
The map on the space of transport coefficient that is induced is given by
\begin{subequations}
\label{example:os}
    \begin{align}
        c_1'&=\qty[c_1-(i(\chi+c_1)\omega-c_2(\omega^2-k^2))\xright_1]_n,\\
        c_2'&=\qty[c_2-(i(\chi+c_1)\omega-c_2(\omega^2-k^2))\xright_2]_{n-1},\\
        c_3'&=\qty[c_3-c_3(i \xright_1 \omega+(\omega^2-k^2)\xright_2]_{n-1},\\
        c_4'&=\qty[c_4+c_3(\xright_1+i \omega \xright_2)]_{n-2}.
    \end{align}
\end{subequations}
In this simple diffusive case, the $K$ matrices from Eqs.~\eqref{eq:rightmult} and \eqref{eq:leftmult} are $1\times 1$, and their product coincides with the polynomial $p$, which is
\begin{equation}
    p(\omega,k,\Xleft,\Xright)=\qty[(1+\xleft_1) (1-i\omega\xright_1+(\omega^2-k^2) \xright_2)]_n.
\end{equation}
Clearly, we have $p(0,0)=1$, as expected. We then have
\begin{equation}
P(\omega,k;\coes')=\qty[P(\omega,k;\coes)p(\omega,k,\Xleft,\Xright)]_{n+1}.
\end{equation}
Similarly, the action on the response function is
\begin{equation}
\label{example:frameaction}
    G^{tt}(\coes')=\frac{\qty[B^{tt}(\omega,k;\coes)p(\omega,k)]_{n+1}}{\qty[P(\omega,k;\coes)p(\omega,k)]_{n+1}}.
\end{equation}
This can be checked explicitly by plugging Eqs.~\eqref{example:frame} and \eqref{example:os} into Eqs.~\eqref{example:P}.

We now expand the coefficients around $\omega=k=0$ to study the modes of the system
\begin{equation}
    c_{i,jl}=\left.\frac{1}{j!l!}(-i\partial_\omega)^j \partial_k^l c_i(\omega,k)\right|_{\omega=k=0}.
\end{equation}
For concreteness, we consider the expansion to second order, i.e., to $n=2$, which gives
\begin{align}
    &\qty[P(\omega,k;\coes)]_{3}=-i\chi \omega+\omega^2(c_{1,10}+c_{2,00})-k^2 c_{2,00} \nonumber \\
    &-ik^2\omega(c_{1,02}+c_{2,10})+i\omega^3(c_{1,20}+c_{2,10}) . 
\end{align}
There is a hydrodynamic (diffusive) mode
\begin{equation}
    \omega_\text{dif}(k)=i\frac{c_{2,00}}{\chi}k^2 +\CO(k^4), \label{example:dif}
\end{equation}
which gives (cf.~Eq.~\eqref{def:CI})
\begin{equation}
\label{eq:invariant}
    \CI(\omega,k;\coes)=\qty[P_\text{hy}(\omega,k;\coes)]_3=\omega-i\frac{c_{2,00}}{\chi}k^2.
\end{equation}
Clearly, $c_{2,00}$ is an invariant, and must be negative to ensure entropy production. This is sensible, since it sets the conductivity (and thereby the diffusion constant), which is a physical and measurable hydrodynamic parameter of a given system. The gapped part of $P$ is (see Eq.~\eqref{def:Pgap})
\begin{align}
    \qty[P_\text{gap}(\omega,k;\coes)]_2=-i\chi+d_1 \omega-id_2 k^2+d_3 \omega^2, \label{example:Phy}
\end{align}
where
\begin{subequations}
\label{example:Pgap}
\begin{align}
    d_1&=c_{1,10}+c_{2,00},\\
    d_2&=c_{1,02}+c_{2,10}-\frac{d_1c_{2,00}}{\chi},\\
    d_3&=c_{1,20}+c_{2,10}.
\end{align}
\end{subequations}
Clearly, there are up to two gapped modes that appear alongside the diffusive mode. It is easily checked that Eqs.~\eqref{example:Pgap} are not invariant under the frame \eqref{example:frame} or on-shell \eqref{example:os} transformations, and may be set to a constant or any other degree-two polynomial via
\begin{equation}
    P_\text{gap}(\omega,k;\coes')=\qty[P_\text{gap}(\omega,k;\coes)p(\omega,k;\Xleft,\Xright)]_2.
\end{equation}

The second invariant, $\mathcal{Y}^{tt}$ (see Eq.~\eqref{def:Jnvariant}) can be expressed as
\begin{align}
    \mathcal{Y}^{tt}(\omega,k;\coes)&=\qty[G^{tt}(\omega,k)P_\text{hy}(\omega,k)]_3\\
    &=-i c_{3,00}k^2+(c_{3,10}+c_{4,00})\omega k^2, \label{example:jnvariant}
\end{align}
with $\mathcal{Y}^{tx}$, $\mathcal{Y}^{xt}$ and $\mathcal{Y}^{xx}$ containing the same information due to Ward identities \eqref{eq:ward}. $\CI(\omega,k;\coes)$ and $\mathcal{Y}^{tt}(\omega,k;\coes)$ contain all the invariant information about this system at $n=2$. Even though, in this case, the invariants are simply the transport coefficients $c_{2,00}$ and $c_{3,00}$, as well as the linear combination $c_{3,10}+c_{4,00}$, the expressions at higher orders are no longer as simple. Due to their considerable length, we do not write their explicit forms here. 

An important point to stress is that either the on-shell transformation or the frame transformations can be chosen to arbitrarily add or remove gapped modes. This means that one make (at least linearised) diffusive hydrodynamics stable and causal in any frame by choosing the appropriate set of on-shell transformations.

Turning to the algebraic structure, it is apparent that the transformation rules \eqref{example:frame} and \eqref{example:os} indeed have the form of \eqref{eq:actionF} and \eqref{eq:actionO}. The algebraic properties cited in Section~\ref{sec:algebra} may be checked explicitly, although we do not do that here. One can verify that the frame and on-shell transformations indeed commute, and, furthermore, have non-zero overlap. For orders $n\leq 3$, the transformations are abelian, with the non-abelian structure appearing at $n=4$. Even for this simple diffusive example, this would require considering $20$ transport coefficients. We defer such analyses to future works. 

We conclude this section by discussing an explicit example, which is accurately described, to all orders, by diffusive hydrodynamics described above. This is the example of the holographic, self-dual linear axion model \cite{Davison:2014lua} (see also Ref.~\cite{Andrade:2013gsa}), describing a finite temperature quantum field theory in three spacetime dimensions, with a conserved U$(1)$ current. The retarded current-current correlator can be computed explicitly in the full microscopic theory, and amounts to (up to some normalisation, which rescales the conserved current $J^\mu$, and thus the equation of state)
\begin{equation}
    \CG^{tt}=\frac{k^2}{2}\frac{\Gamma\qty(\frac{1}{4}-i\frac{\omega}{2}-\frac{1}{4}\Delta(k))\Gamma\qty(\frac{1}{4}-i\frac{\omega}{2}+\frac{1}{4}\Delta(k))}{\Gamma\qty(\frac{3}{4}-i\frac{\omega}{2}-\frac{1}{4}\Delta(k))\Gamma\qty(\frac{3}{4}-i\frac{\omega}{2}+\frac{1}{4}\Delta(k))},
\end{equation}
where $\Delta(k)=\sqrt{1-4k^2}$. Here, temperature is set to $T=1/2\pi$. In these units, the susceptibility of this model is $ \chi=1$. To construct a hydrodynamic approximation to the exact response function, we can rewrite $\CG^{tt}$ as
\begin{equation}
    \CG^{tt}(\omega,k)=\frac{\CB^{tt}(\omega,k)}{\CP(\omega,k)},
\end{equation}
where
\begin{subequations}
\label{example:BP}
\begin{align}
    \CB^{tt}(\omega,k)&=-i\, \mathcal{Y}^{tt}(\omega,k) p(\omega,k),\\
    \CP(\omega,k)&=-i\, \CI(\omega,k) p(\omega,k).
\end{align}
\end{subequations}
Here, 
\begin{subequations}
\begin{align}
    \CI(\omega,k)&=\omega-i\frac{\Delta(k)-1}{2},\\
    \mathcal{Y}^{tt}(\omega,k)&=i k^2\frac{\Gamma\qty(\frac{5}{4}-i\frac{\omega}{2}-\frac{1}{4}\Delta(k))\Gamma\qty(\frac{1}{4}-i\frac{\omega}{2}+\frac{1}{4}\Delta(k))}{\Gamma\qty(\frac{3}{4}-i\frac{\omega}{2}-\frac{1}{4}\Delta(k))\Gamma\qty(\frac{3}{4}-i\frac{\omega}{2}+\frac{1}{4}\Delta(k))},
\end{align}
\end{subequations}
and $p(\omega,k)$ can be an arbitrary function with $p(0,0)=1$, which is also analytic around $\omega=k=0$. Performing a hydrodynamic expansion in the sense of Eq.~\eqref{eq:hydro-approx} gives the truncated hydrodynamic response function
\begin{equation}
    G^{tt}(\omega,k)=\frac{\qty[\CB^{tt}(\omega,k)]_{n+1}}{\qty[\CP(\omega,k)]_{n+1}}.
\end{equation}
Clearly, the function $p(\omega,k)$ in Eqs.~\eqref{example:BP} encapsulates the frame/on-shell ambiguity of Eq.~\eqref{example:frameaction}. We can now match the invariant polynomials $\CI$ and $\mathcal{Y}^{tt}$ to the exact response functions. Up to second order in the hydrodynamic expansion, we have
\begin{subequations}
\label{example:IY}
\begin{align}
    \CI(\omega)&=\omega+i k^2+\ldots,\\
    \mathcal{Y}^{tt}(\omega)&=ik^2+\ldots.
\end{align}    
Matching this with Eqs.~\eqref{eq:invariant} and \eqref{example:jnvariant} amounts to matching the hydrodynamic effective description to the exact microscopic theory.
\end{subequations}

\section{Discussion and future directions}
In this paper, we systematically treated, in the most general sense, linearised, classical hydrodynamics to arbitrary order in derivative expansion. This allowed us to discuss the physical excitations (the modes) of hydrodynamic states, as well as the hydrodynamic retarded two-point correlation or response functions. We explained how such response functions can be expanded systematically in a way that is consistent with the gradient expansion. By building on this formalism, we then treated, in detail, the frame and on-shell transformations as maps in the space of transport coefficients. Their action on the hydrodynamic response functions was shown to be universal and independent of the specifics of the theory. From a more formal point of view, we also constructed (algebraic) invariants, among which are the small-$k$ expansions of the hydrodynamic dispersion relations. Finally, we discussed several group theoretic properties of the frame and on-shell transformations, among which is their nilpotent group structure.

Even though we limited ourselves to the discussion of linearised hydrodynamics, we expect that a large part of the present analysis should be applicable to the full, non-linear theory. One way to understand this is at the level of equations of motion, where the frame and on-shell transformations acting on the space of transport coefficients should again provide equivalence relations between different effective theories with a single underlying microscopic description. While the frame transformations have been studied extensively in this context, especially with regards to constructing a stable and causal theory of relativistic hydrodynamics, the on-shell transformations, which behave in a similar way, have not been explored in this manner. Since there exist on-shell transformations that do not affect the frame choice, but nevertheless induce transformations on transport coefficients, it is possible that a causal theory of relativistic hydrodynamics may be constructed even in the Landau frame. Another way would be to consider higher-point correlation functions, where the low-$\omega$ and $k$ parts of their spectra are again expected to correspond to certain (algebraic) invariants. Furthermore, it would be interesting to pursue similar explorations from the point of view of the full hydrodynamic EFT with stochastic and quantum fluctuations, which is known to introduce (infinitely many) branch cuts into correlation functions \cite{Chen-Lin:2018kfl,Grozdanov:2024fle}. Another interesting set of questions pertains to the discussion of (quasi)-hydrodynamic theories with weakly broken symmetries \cite{Grozdanov:2018fic}, where the hydrodynamic modes acquire a parametrically small gap. Lastly, it would be interesting to consider the possibility of formulating hydrodynamics purely in terms of the manifestly invariant dynamical quantities.

Beyond hydrodynamics, many other interesting EFTs exist in the literature, which are also derivative expansions. Examples include the chiral effective theory, the Euler-Heisenberg Lagrangian and higher-derivative theories of gravity (or supergravity) that can be systematically derived from string theory. The history of discussions on the field-redefinitions (often higher-derivative) in such theories is long (see e.g.~Refs.~\cite{Metsaev:1987ju,Tseytlin:1993df,Grozdanov:2014kva,Liu:2022sew}). One may wonder if the methods developed here could be used in systematising of the field-redefinitions in those theories.

\begin{acknowledgments}
We thank Sayantani Bhattacharyya, Pavel Kovtun and Geoff Vasil for illuminating discussions. The work of S.G. was supported by the STFC Ernest Rutherford Fellowship ST/T00388X/1. The work is also supported by the research programme P1-0402 and the project J7-60121 of Slovenian Research Agency (ARIS). M.V. is funded by the STFC Studentship ST/X508366/1.
\end{acknowledgments}

\onecolumngrid
\appendix
\section{Systematics of series expansions}
\label{app:series}
We begin by stating some useful properties of series expansions for functions of two variables. They all follow from the basic properties of multivariate Taylor expansion. $f(\omega,k)$ and $g(\omega,k)$ denote non-zero functions that are analytic around $\omega=k=0$ throughout. We define the following:
\begin{itemize}
    \item $\CO_n(\omega,k)$ is any function with $\CO_n(\lambda\omega,\lambda k)=\CO(\lambda^n)$. This means that we have
    \begin{equation}
        g(\omega,k)=f(\omega,k)+\CO_{n+1}(\omega,k)\Leftrightarrow \qty[g(\omega,k)]_n=\qty[f(\omega,k)]_n.
    \end{equation}
    \item The \emph{degree} of a polynomial $f(\omega,k)$ is the lowest integer $M_f$ for which $f(\omega,k)=\qty[f(\omega,k)]_{M_f}$.
    \item The \emph{lower degree} of a polynomial $f(\omega,k)$ is the largest integer $m_f$ for which $\qty[f(\omega,k)]_{m_f-1}=0$.
    \item A polynomial $f(\omega,k)$ is \emph{homogeneous of degree} $N$ if $M_f=m_f=N$, which means that
    \begin{equation}
        f(\lambda\omega,\lambda k)=\lambda^N f(\omega,k).
    \end{equation}
\end{itemize}
We will use the following identities throughout:
\begin{itemize}
    \item If $m_f$ and $m_g$ are the lower degrees of $f(\omega,k)$ and $g(\omega,k)$ respectively, then
    \begin{equation}
    \qty[f(\omega,k)g(\omega,k)]_n=\qty[\qty[f(\omega,k)]_{n-m_g}\qty[g(\omega,k)]_{n-m_f}]_n \label{id:product}
    \end{equation}
    \item If $f(\omega,k)$ is homogeneous of degree $N$, then
    \begin{equation}    f(\omega,k)\qty[g(\omega,k)]_n=\qty[f(\omega,k)g(\omega,k)]_{N+n}. \label{id:hom}
    \end{equation}
\end{itemize}

\section{Constitutive relations and response functions}

\label{app:correlators}
By splitting the conserved currents into their equilibrium and perturbed parts
\begin{equation}
    (\CJ_{(i)})^\mu=(\CJ^0_{(i)})^\mu+(\delta\CJ_{(i)})^\mu,
\end{equation}
we can write the divergence of a current as
\begin{equation}
    (\delta \CJ_{(i)})^\mu \xrightarrow{\text{div}} i k_\mu (\delta\CJ_{(i)})^\mu,
\end{equation}
where $k_\mu=(-\omega,\vb{k})$. By reorganising all the independent components of conserved currents into a $M$-dimensional tuple $\CJ^a$, the divergence is described by a matrix $q^A{}_b$ as
\begin{equation}
    \delta \CJ^a \xrightarrow{\text{div}} q^A{}_b(\omega,k) \delta \CJ^b. \label{app:qdef}
\end{equation}
Here, $q^A{}_{b}(\omega,k)$ is homogeneous of degree one, as its elements must be linear in $\omega$ and $\vb k$. It maps from the $M$-dimensional space of the independent components of conserved currents to the $N$-dimensional space of independent conservation equations. The latter can be rewritten as (see Eq.~\eqref{eq:conservation})
\begin{equation}
    q^A{}_b(\omega,k)\delta \CJ^b=\CS^{Ab}(\omega,k) \delta \CA_b.
\end{equation}
The (retarded) linear response of conserved currents to the perturbation of external sources in encapsulated in the full microscopic response function
\begin{equation}
    \delta \CJ^b=\CG^{ab}(\omega,k)\delta \CA_b.
\end{equation}
The components of the response function are not independent as they obey the Ward identities
\begin{equation}
\label{app:ward}
    q^A{}_a(\omega,k)\CG^{ab}(\omega,k)=\CS^{Ab}(\omega,k).
\end{equation}

By the hydrodynamic assumption, all conserved currents are functions of the $N$ hydrodynamic variables $\delta\mu_A$ and $M$ external sources $\delta \CA_a$. This means that we can write
\begin{equation}
\label{app:const}
    \delta\CJ^a(\omega,k;\delta\mu,\delta\CA;\coes)=\CJ^{aB}(\omega,k;\coes)\delta\mu_B+\CJ^{ab}(\omega,k;\coes)\delta \CA_b.
\end{equation}
In full, non-linear $n$-th order hydrodynamics, constitutive relations amount to summing over all the tensors which are functions $\mu$ and $\CA$, are at least first and at most $n$-th order in derivatives, and are compatible with the symmetries and transformation rules of conserved currents. By increasing the order $n$, the number of such tensors generally increases indefinitely. When considering linearised hydrodynamics, however, many such tensors are linearly dependent, and related to one another via functions of $\omega$ and $k$. The linearised constitutive relations then take the form (see Eqs.~\eqref{def:Jcurrent} and \eqref{eq:constitutive})
\begin{align}
    \CJ^{aB}(\omega,k;\coes)&=\qty[\CJ^{aB}_\text{ideal}+\sum_\fati c_\fati(\omega,k) \CJ^{aB}_\fati(\omega,k)]_n+\CO_{n+1}(\omega,k),\\
    \CJ^{ab}(\omega,k;\coes)&=\qty[\CJ^{ab}_\text{ideal}+\sum_\fati c_\fati(\omega,k) \CJ^{ab}_\fati(\omega,k)]_n+\CO_{n+1}(\omega,k).
\end{align}
Here, $\CJ_\fati^{aB}$ and $\CJ_\fati^{ab}$ are induced through $\CJ^a_\fati=\CJ_\fati^{aB}\delta\mu_B+\CJ_\fati^{ab}\delta \CA_b$. We pick the basis of $\CJ_\fati^a$ so that they are homogeneous of degree $n_\fati$. Out of the infinite number of such choices, we pick the one for which $n_\fati$ are minimal. To ensure the vanishing of the dissipative part of the constitutive relations in equilibrium, we must either have $n_\fati\geq 1$, or $c_\fati(0,0)=0$. In other words, the lower degree of $c_\fati(\omega,k) \CJ_\fati^a(\omega,k)$ must be at least one (see the $c_1(\omega,k)$ in the constitutive relations \eqref{example:constitutive2}). This truncates the coefficient functions $c_\fati(\omega,k)$ at $\qty[c_\fati(\omega,k)]_{n-n_\fati}$. Although we are truncating the derivative expansion at $n$-th order, we are systematically tracking the influence of higher-order terms through $\CO_{n+1}(\omega,k)$.

The equations of motion arise from conservation equations, and take the form
\begin{equation}
\label{app:eom}
    F^{AB}(\omega,k;\coes)\delta\mu_B=Q^{Ab}(\omega,k;\coes)\delta\CA_b,
\end{equation}
where
\begin{subequations}
\begin{align}
    F^{AB}(\omega,k;\coes)&=q^A{}_c(\omega,k) \CJ^{cB}(\omega,k;\coes)+\CO_{n+2}(\omega,k),\\
    Q^{Ab}(\omega,k;\coes)&=S^{Ab}(\omega,k)-q^{A}{}_c(\omega,k)\CJ^{cb}(\omega,k;\coes)+\CO_{n+2}(\omega,k).
\end{align}    
\end{subequations}
Motivated by concrete examples, we assume that $\CS^{Ab}$ is homogeneous of degree one.

To find the response functions, we first express the solutions of \eqref{app:eom} as
\begin{equation}
    \delta\mu_B=(F^{-1})_{BA}Q^{Ab}\delta \CA_b.
\end{equation}
Plugging it into the constitutive relations \eqref{app:const}, we obtain
\begin{equation}
    \delta \CJ^a(\omega,k;F^{-1}Q\delta\CA,\delta\CA;\coes)=G_\text{full}^{ab}(\omega,k;\coes)\delta \CA_b,
\end{equation}
with
\begin{equation}
G_\text{full}^{ab}(\omega,k;\coes)=\frac{B_\text{full}^{ab}(\omega,k;\coes)+\CO_{N+n+1}(\omega,k)}{\det F(\omega,k;\coes)+\CO_{N+n+1}(\omega,k)}. \label{app:G-full}
\end{equation}
Here, $B_\text{full}^{ad}$ is a polynomial in $\omega$ and $k$ with explicit form
\begin{equation}
\label{app:Bexpression}
        B_\text{full}^{ad}={\CJ^{aB}(F^\ddagger)_{BC}Q^{Cb}}+\CJ^{ab}\det F.
\end{equation}
$F^\ddagger$ is the adjugate matrix of $F$, i.e., the transpose of its cofactor matrix, for which the following holds:
\begin{equation}
    F^\ddagger=F^{-1}\det F.
\end{equation}
The virtue of the adjugate matrix is the fact that it is polynomial in the elements of $F$. Expression \eqref{app:G-full} is the generic form of a response function in any classical hydrodynamic theory. Note that, in general, $B^{ab}_\text{full}(\omega,k;\coes)$ and $\det F(\omega,k;\coes)$ are polynomials of degree $N(n+1)$. However, as indicated in Eq.~\eqref{app:G-full}, higher orders are sensitive to the $\CO_{n+1}(\omega,k)$ term in the constitutive relations. This ambiguity is eliminated by introducing the truncated response function
\begin{equation}
    G^{ab}(\omega,k;\coes)=\frac{B^{ab}(\omega,k;\coes)}{P(\omega,k;\coes)}=\frac{\qty[B_\text{full}^{ab}(\omega,k;\coes)]_{N+n+1}}{\qty[\det F(\omega,k;\coes)]_{N+n}}.
\end{equation}
Putting the above assumptions and results together, we can conclude that $B^{ab}(\omega,k;\coes)$ and $P(\omega,k;\coes)$ are both of lower degree $N$.

\section{Frame transformations}
\label{app:frame}
The most general redefinition of the hydrodynamic variables that preserves their equilibrium value is written as
\begin{equation}
    \delta\mu_A(\delta\mu')=\delta\mu'_A+\sum_\fatj {\xleft}_\fatj(\omega,k) \CK_{\fatj,A} (\omega,k;\delta\mu',\delta\CA) +\CO_{n+1}(\omega,k). \label{app:frameEQ}
\end{equation}
We pick the basis of $\CK_{\fatj,A}$ so that they are homogenous of degree $\cev q_\fatj$. Out of the infinite number of such choices, we pick the one for which $\cev q_\fatj$ are minimal. To ensure the consistency of the equilibrium values of the hydrodynamic variables, we must either have $\cev q_\fatj\geq1$ or $\xleft_\fatj(0,0)=0$. Eq.~\eqref{app:frameEQ} can be rewritten as
\begin{equation}
    \delta\mu_A=\Kleft_A{}^B(\omega,k;\Xleft) \delta\mu'_B+\Omega_A{}^b(\omega,k;\Xleft)\delta\CA_b+\CO_{n+1}(\omega,k),
\end{equation}
where
\begin{subequations}
\label{app:KOmegadef}
    \begin{align}
        \Kleft_A{}^B(\omega,k;\Xleft)&=\delta_A^B+\sum_{\fatj} \xleft_{\fatj}(\omega,k)\CK_{\fatj,A}{}^B(\omega,k),\\
        \Omega_A{}^b(\omega,k;\Xleft)&=\sum_{\fatj}\xleft_{\fatj}(\omega,k)\CK_{\fatj,A}{}^b(\omega,k).
    \end{align}
\end{subequations}
Here, $\CK_{\fatj,A}{}^B$ and $\CK_{\fatj,A}{}^b$ are induced through $\CK_{\fatj,A}=\CK_{\fatj,A}{}^B\delta\mu_B+\CK_{\fatj,A}^b\delta\CA_b$. A frame transformation is defined on $\coes$ as (see Eq.~\eqref{def:framemap})
\begin{subequations}
\label{app:fr:Jaction}
\begin{align}
    \CJ^{aB}(\omega,k;\coes')&=\qty[\CJ^{aA}(\omega,k;\coes) \Kleft_{A}{}^B(\omega,k;\Xleft)]_n +\CO_{n+1}(\omega,k),\\
    \CJ^{ab}(\omega,k;\coes')&=\qty[\CJ^{ab}(\omega,k;\coes')+\CJ^{aA}(\omega,k;\coes)\Omega_{A}{}^b(\omega,k;\Xleft)]_n+\CO_{n+1}(\omega,k).
\end{align}    
\end{subequations}
Note that this truncates the frame coefficient functions $\xleft_\fatj(\omega,k)$ at $\qty[\xleft_\fatj(\omega,k)]_{n-\cev q_\fatj}$. Taking the divergence of the above, we get
\begin{subequations}

\begin{align}
    F^{AB}(\omega,k;\coes')&=\qty[F^{AB}(\omega,k;\coes)\Kleft_A{}^B(\omega,k;\Xleft)]_{n+1}+\CO_{n+2}(\omega,k),\\
    Q^{Ab}(\omega,k;\coes')&=\qty[Q^{Ab}(\omega,k;\coes)-F^{AB}(\omega,k;\coes)\Omega_B{}^b(\omega,k;\Xleft)]_{n+1}\CO_{n+2}(\omega,k).
\end{align}    
\end{subequations}
Using the identities from Appendix \ref{app:series}, we get, after some lines of simple manipulation,
\begin{subequations}
\begin{align}
    \det F(\omega,k;\coes')&=\det F(\omega,k;\coes) \det \Kleft(\omega,k;\Xleft)+\CO_{N+n+1}(\omega,k),\\
    B_\text{full}^{ab}(\omega,k;\coes')&=B_\text{full}^{ab}(\omega,k;\coes) \det \Kleft(\omega,k;\Xleft)+\CO_{N+n+1}(\omega,k).
\end{align}    
\end{subequations}
This means that the action on the truncated quantities is simple
\begin{subequations}
\begin{align}  P(\omega,k;\coes')&=\qty[P(\omega,k;\coes)\det\Kleft(\omega,k;\Xleft)]_{N+n}, \label{app:Pframe}\\
B^{ab}(\omega,k;\coes')&=\qty[B^{ab} (\omega,k;\coes)\det\Kleft(\omega,k;\Xleft)]_{N+n}.
\end{align}
    
\end{subequations}

Finally, we treat the action of frame transformations on coefficients themselves. Firstly, we note that the basis of $\CK_\fatj$ is assumed to be, in the appropriate sense, complete. This means that there exists $\rholeft_{\fati \fatj \fatk}(\omega,k)$ such that
\begin{subequations}
\label{app:fr:kalgebra}
    \begin{align}
        \CK_{\fati,A}{}^B\CK_{\fatj,B}{}^C&=\sum_{\fatk}\rholeft_{\fatj\fati\fatk} \CK_{\fatk,A}{}^C,\\
        \CK_{\fati,A}{}^B\CK_{\fatj,B}{}^c&=\sum_{\fatk}\rholeft_{\fatj\fati\fatk}\CK_{\fatk,A}{}^c.
    \end{align}
\end{subequations}
Furthermore, the basis of $\CJ_\fati$ is also assumed to be, in the appropriate sense, complete. This means that there exists $(\gammaleft_\fati)_{\fatj}(\omega,k)$, such that
\begin{subequations}
\label{app:fr:gammaalgebra}
\begin{align}
    \CJ_\text{ideal}^{aB}\CK_{\fati,B}{}^C&=\sum_{\fatj}(\gammaleft_{\fati})_\fatj\CJ_{\fatj}^{aC},\\ \CJ_{\text{ideal}}^{aB}\CK_{\fati,B}{}^c&=\sum_{\fatj}(\gammaleft_{\fati})_{\fatj}\CJ_{\fatj}^{ac},
\end{align}    
\end{subequations}
and $(\sigmaleft_\fati)_{\fatj\fatk}(\omega,k)$, such that
\begin{subequations}
\label{app:fr:sigmaalgebra}
\begin{align}
    \CJ_{\fatk }^{aB}\CK_{\fati,B}{}^C&=\sum_{\fatj}(\sigmaleft_{\fati})_{\fatj\fatk}\CJ_{\fatj}^{aC},\\ \CJ_{\fatk}^{aB}\CK_{\fati,B}{}^c&=\sum_{\fatj}(\sigmaleft_{\fati})_{\fatj\fatk}\CJ_{\fatj}^{ac}.
\end{align}    
\end{subequations}
The explicit form of the frame transformation in the space of transport coefficients is
\begin{equation}
\label{app:coesF-action}
c_{\fatj}'(\omega,k)=\qty[c_{\fatj}+\sum_{\fati}\xleft_{\fati}\qty((\gammaleft_{\fati})_{\fatj}+\sum_{\fatk}(\sigmaleft_{\fati})_{\fatj\fatk}c_{\fatk})]_{n-n_\fatj}.
\end{equation}
The vectors $(\gammaleft_\fati)_\fatj$ and the matrices $(\sigmaleft_\fati)_{\fatj\fatk}$ are endowed with structure. From matrix associativity and Eqs.~\eqref{app:fr:kalgebra}, \eqref{app:fr:gammaalgebra}, and \eqref{app:fr:sigmaalgebra}, it follows that
\begin{subequations}
\label{app:sigmaleftalgebra}
\begin{align}
    \sigmaleft_\fati \sigmaleft_\fatj &= \sum_\fatk\rholeft_{\fati\fatj\fatk}\sigmaleft_\fatk,\\
    \sigmaleft_\fati \gammaleft_\fatj &= \sum_\fatk\rholeft_{\fati\fatj\fatk}\gammaleft_\fatk.
\end{align}    
\end{subequations}
That is, they form an algebra. Matrices $\sigmaleft_\fati$ obey the same algebra for right multiplication as the matrices $\CK_\fati$ obey for the left multiplication (see Eq.~\eqref{app:fr:Jaction}). Furthermore, if we introduce matrices $\rholeft_\fati$ as $(\rholeft_\fati)_{\fatj\fatk}=\rholeft_{\fatj\fati\fatk}$, we get
\begin{equation}
\label{app:rholeftalgebra}
    \rholeft_\fati \rholeft_\fatj = \sum_\fatk\rholeft_{\fati \fatj \fatk}\rholeft_\fatk,
\end{equation}
which follows from matrix associativity.

\section{On-shell transformations}
\label{app:os}
To treat the on-shell transformations, we introduce $\delta\xi_{(i)}(\coes)$ as in Eq.~\eqref{def:xi}. We also introduce $\xi^A(\omega,k;\coes)$, which is an $N$-dimensional tuple of all the $\delta\xi_{(i)}(\coes)$ so that
\begin{equation}
    \xi^A(\omega,k;\coes)=F^{AB}(\omega,k;\coes)\delta\mu_B-Q^{Ab}(\omega,k;\coes)\delta \CA_b.
\end{equation}
The on-shell transformation is then defined as (see Eq.~\eqref{def:os})
\begin{equation}
    \delta \CJ^a(\omega,k;\coes')=\qty[\delta \CJ^a(\omega,k;\coes)+\sum_{\fatj}\xright_{\fatj}(\omega,k)\CH_{\fatj}^a{}_{B}(\omega,k)\xi^B(\omega,k;\coes)]_n+\CO_{n+1}(\omega,k).
\end{equation}
Here, $\CH_\fatj^a{}_B$ is induced through $\CH_\fatj^a=\CH_\fatj^a{}_B\xi^B$. We pick the basis of $\CH_\fatj^a{}_B$ so that they are homogeneous of degree $\vec q_{\fatj}-1$, where the offset of one is introduced because they appear with the divergence matrix $q^A{}_b$ multiplying them from the right. The on-shell transformation truncates the on-shell coefficient functions $\xright_\fatj(\omega,k)$ at $\qty[\xright_{\fatj}(\omega,k)]_{n-\vec q_{\fatj}}$. By introducing
\begin{subequations}
\begin{align}
    W^{a}{}_B(\omega,k;\Xright)&=\sum_{\fatj}\xright_{\fatj}(\omega,k)\CH_{\fatj}^a{}_{B}(\omega,k) \label{app:Wdef},\\
    \Kright^{A}{}_B(\omega,k;\Xright)&=\delta^A_B+q^A{}_b(\omega,k)W^{b}{}_B(\omega,k;\Xright),
\end{align}    
\end{subequations}
we obtain the transformation rules
\begin{subequations}
\label{app:OSdef}
\begin{align}
    \CJ^{aB}(\omega,k;\coes')&=\qty[\CJ^{aB}(\omega,k;\coes)+W^{a}{}_A(\omega,k;\Xright) F^{AB}(\omega,k;\coes)]_n+\CO_{n+1}(\omega,k),\\
    \CJ^{ab}(\omega,k;\coes')&=\qty[\CJ^{ab}(\omega,k;\coes)-W^a{}_{B}(\omega,k;\Xright) Q^{Bb}(\omega,k;\coes)]_n+\CO_{n+1}(\omega,k).
\end{align}    
\end{subequations}
Taking the divergence, we then get
\begin{subequations}
\begin{align}
    F^{AB}(\omega,k;\coes')&=\qty[\Kright^A{}_C(\omega,k;\Xright) F^{CB}(\omega,k;\coes)]_{n+1}+\CO_{n+2}(\omega,k),\\
    Q^{Ab}(\omega,k;\coes')&=\qty[\Kright^A{}_B(\omega,k;\Xright) Q^{Bb}(\omega,k;\coes)]_{n+1}+\CO_{n+2}(\omega,k).
\end{align}    
\end{subequations}
Using the identities from Appendix \ref{app:series}, we finally find
\begin{subequations}
\begin{align}
    \det F(\omega,k;\coes')&=\det F(\omega,k;\coes) \det \Kright(\omega,k;\Xright)+\CO_{N+n+1}(\omega,k),\\
    B^{ab}(\omega,k;\coes')&=B^{ab}(\omega,k;\coes)\det\Kright(\omega,k;\Xright)+\CO_{N+n+1}(\omega,k).
\end{align}    
\end{subequations}
The action on the truncated quantities is therefore simple:
\begin{subequations}
\begin{align}  P(\omega,k;\coes')&=\qty[P(\omega,k;\coes)\det\Kright(\omega,k;\Xright)]_{N+n}\label{def:Pshell},\\
B^{ab}(\omega,k;\coes')&=\qty[B^{ab}(\omega,k;\coes)\det\Kright(\omega,k;\Xright)]_{N+n}.
\end{align}

\end{subequations}

Finally, we consider the action of the on-shell transformations in the space of transport coefficients. Because the basis of $\CH_\fatj$ is assumed to be, in the appropriate sense, complete, there must exist $\rhoright_{\fati\fatj\fatk}(\omega,k)$, such that
\begin{equation}
\label{app:os:rhodef}
    \CH_\fati^a{}_B q^B{}_c\CH_\fatj^c{}_C=\sum_\fatj \rhoright_{\fati\fatj\fatk} \CH_\fatk^a{}_C.
\end{equation}
Note that the coefficients $\rhoright_{\fati\fatj\fatk}$ are defined differently than in the case of frame transformations \eqref{app:fr:kalgebra}. This is because their natural action is, in contrast with the frame case, as matrix multiplication from the left. Furthermore, because basis of $\CJ_\fati$ is also assumed to be, in the appropriate sense, complete. This means that there exists $(\gammaright_\fati)_{\fatj}(\omega,k)$ such that
\begin{subequations}
\label{app:os:gammadef}
\begin{align}
    \CH_\fati^a{}_B q^B{}_b\CJ_\text{ideal}^{bC}&=\sum_\fatj (\gammaright_\fati)_\fatj \CJ^{aC}_\fatj,\\
    \CH_\fati^a{}_B \qty(q^B{}_b\CJ_\text{ideal}^{bc}- S^{Bc})&=\sum_\fatj (\gammaright_\fati)_\fatj \CJ^{ac}_\fatj,
\end{align}
\end{subequations}
and $(\sigmaright_\fati)_{\fatj\fatk}(\omega,k)$, such that
\begin{subequations}
\label{app:os:sigmaalgebra}
\begin{align}
    \CH_\fati^a{}_B q^B{}_b \CJ_\fatk^{bC}&=\sum_\fatj (\sigmaright_\fati)_{\fatj\fatk}\CJ_\fatj^{aC},\\
    \CH_\fati^a{}_B q^B{}_b \CJ_\fatk^{bc}&=\sum_\fatj (\sigmaright_\fati)_{\fatj\fatk}\CJ_\fatj^{ac}.
\end{align}
\end{subequations}
The explicit action on transport coefficients is given by
\begin{equation}
\label{app:coesO-action}
c_{\fatj}'(\omega,k)=\qty[c_{\fatj}+\sum_{\fati}\xright_{\fati}\qty((\gammaright_{\fati})_{\fatj}+\sum_{\fatk}(\sigmaright_{\fati})_{\fatj\fatk}c_{\fatj})]_{n-n_\fatj}.
\end{equation}
Analogously to the frame case, $(\gammaright_\fati)_\fatj$ and $(\sigmaright_\fati)_{\fatj\fatk}$ are endowed with structure. From matrix associativity and Eqs.~\eqref{app:os:rhodef}, \eqref{app:os:gammadef} and $\eqref{app:os:sigmaalgebra}$, it follows that
\begin{subequations}
\label{app:sigmarightalgebra}
\begin{align}
    \sigmaright_\fati \sigmaright_\fatj &= \sum_\fatk\rhoright_{\fati\fatj\fatk}\sigmaright_\fatk,\\
    \sigmaright_\fati \gammaright_\fatj &= \sum_\fatk\rhoright_{\fati\fatj\fatk}\gammaright_\fatk,
\end{align}    
\end{subequations}
i.e., they form an algebra. If we introduce matrices $\rhoright_\fati$ as $(\rhoright_\fati)_{\fatj\fatk}=\rhoright_{\fatj\fati\fatk}$, we get
\begin{equation}
\label{app:rhorightalgebra}
    \rhoright_\fati \rhoright_\fatj = \sum_\fatk\rhoright_{\fati \fatj \fatk}\rhoright_\fatk,
\end{equation}
which is completely analogous to the discussion about frame transformations.

\section{Invariants}
\label{app:invariants}
The goal of this appendix is to show the invariance of the polynomials $\CI(\omega,k;\coes)$ and $\mathcal{Y}^{ab}(\omega,k;\coes)$, as defined in Eqs.~\eqref{def:CI} and \eqref{def:Jnvariant}.

We begin by stating an identity which will prove to be useful. Let $f(\omega,k)$ be analytic around $\omega=k=0$, and of lower degree $N$. Let $\omega(k)$ be a function that is analytic around $k=0$. The following then holds:
\begin{equation}
f\qty(\omega(k)+\CO(k^{n+2}),k)=f\qty(\omega(k),k)+
\begin{cases}
    \CO(k^{n+2}), &\omega(0)\neq0,\\
    \CO(k^{N+n+1}). &\omega(0)=0.
\end{cases}
 \label{app:megaidentity}
\end{equation}
This is simple to prove through basic properties of series expansions. We emphasise that we use the identity \eqref{id:product} throughout this appendix.

We now apply this statement to the determinant $\det F(\omega,k;\coes)$. By construction, $F$ is a $N\times N$ matrix that is acquired by taking $N$ divergences. Its determinant is therefore necessarily of lower order $N$, with
\begin{equation}
    \qty[\det F(\omega,0;\coes)]_N\neq 0.
\end{equation}
This property gives rise to $N$ hydrodynamic modes $\widetilde\omega_i(k;\coes)$ \cite{Grozdanov:2019uhi}, obeying
\begin{equation}
    \det F(\widetilde\omega_i(k;\coes),k;\coes)=0, \quad \widetilde\omega_i(0;\coes)=0, \quad i=1,\ldots,N.
\end{equation}
By definition of $P(\omega,k;\coes)$, we have
\begin{equation}
    P(\omega,k;\coes)=\det F(\omega,k;\coes)+\CO_{N+n+1}(\omega,k), \label{app:PF}
\end{equation}
with respective hydrodynamic modes, denoted by $\omega_i(k)$, obeying
\begin{equation}
    P(\omega_i(k;\coes),k;\coes)=0, \quad \omega_i(0;\coes)=0, \quad i=1,,\ldots,N.
\end{equation}
Plugging $\omega_i(k;\coes)$ into Eq.~\eqref{app:PF} gives
\begin{equation}
    \det F(\omega_i(k;\coes),k;\coes)=\CO_{N+n+1}(\omega_i(k;\coes),k)=\CO(k^{N+n+1}). \label{app:modeEQ}
\end{equation}
Note that in the case of a gapped mode, we would have $\CO(k^0)$ on the right hand side. By comparing Eq.~\eqref{app:modeEQ} with Eq.~\eqref{app:megaidentity}, we can conclude
\begin{equation}
    \qty[\omega_i(k;\coes)]_{n+1}=\qty[\widetilde\omega_i(k;\coes)]_{n+1}.
\end{equation}
That is precisely the reliable part of the $k$-expansion of the dispersion relations (see Eq.~\eqref{def:varpi}). We now show that it is invariant under frame and on-shell transformations. The most general such transformations act on $P(\omega,k;\coes)$ as (see Eq.~\eqref{def:p} and Eqs.~\eqref{app:Pframe} and \eqref{def:Pshell})
\begin{equation}
    P(\omega,k;\coes')=P(\omega,k;\coes)p(\omega,k;\Xnone)+\CO_{N+n+1}(\omega,k) .\label{app:Ptrans}
\end{equation}
Plugging $\omega_i(k,\coes)$ into Eq.~\eqref{app:Ptrans}, we get
\begin{equation}
    P(\omega_i(k;\coes),k;\coes')=\CO_{N+n+1}(\omega_i(k;\coes),k)=\CO(k^{N+n+1}). \label{app:framehy}
\end{equation}
Again, we note that the right hand side would be $\CO(k^0)$ for gapped modes.
By comparing Eq.~\eqref{app:framehy} with Eq.~\eqref{app:megaidentity}, we can conclude
\begin{equation}
    \qty[\omega_i(k;\coes)]_{n+1}=\qty[\omega_i(k;\coes')]_{n+1}.
\end{equation}
It is therefore clear that $\qty[\omega_i(k;\coes)]_{n+1}$, and therefore the dispersion relation coefficients $\varpi_{ij}$ defined in Eq.~\eqref{def:varpi} are invariants.

We can now split $P(\omega,k;\coes)$ into the hydrodynamic and the gapped parts
\begin{equation}
    P(\omega,k;\coes)=P_\text{hy}(\omega,k;\coes) P_\text{gap}(\omega,k;\coes),
\end{equation}
where
\begin{equation}
    P_\text{hy}(\omega,k;\coes)\equiv\prod_{i=1}^N (\omega-\omega_i(k;\coes)).
\end{equation}
Since $P_\text{hy}$ is of lower degree $N$, we can write (see Eq.~\eqref{id:product})
\begin{equation}
    P(\omega,k;\coes)=\qty[\qty[P_\text{hy}(\omega,k;\coes)]_{N+n}\qty[P_\text{gap}(\omega,k;\coes)]_n]_{N+n},
\end{equation}
meaning that
\begin{equation}
    \CI(\omega,k;\coes)\equiv\qty[P_\text{hy}(\omega,k;\coes)]_{N+n}
\end{equation}
is an invariant polynomial, and that the transformation \eqref{app:Ptrans} acts as
\begin{equation}
    P_\text{gap}(\omega,k;\coes')=\qty[P_\text{gap}(\omega,k;\coes)p(\omega,k;\Xnone)]_n.
\end{equation}
It is then straight-forward to infer from Eq.~\eqref{id:product} that we can engineer $p(\omega,k;\Xleft,\Xright)$ to add or remove gapped modes. Finally, since 
\begin{equation}
    B^{ab}(\omega,k;\coes')=\qty[B^{ab}(\omega,k;\coes)p(\omega,k;\Xnone)]_{N+n},
\end{equation}
we can conclude that
\begin{equation}
    \mathcal{Y}^{ab}(\omega,k;\coes)\equiv\qty[\frac{B^{ab}(\omega,k;\coes)}{P_\text{gap}(\omega,k;\coes)}]_{N+n}
\end{equation}
is an invariant polynomial. This is equivalent to the definition \eqref{def:Jnvariant}. Here, we have used the fact that the components of $B^{ab}$ are all functions of lower degree $N$, which can be inferred from the expression \eqref{app:Bexpression}. It follows from Eq.~\eqref{app:ward} that the invariants $\mathcal{Y}^{ab}$ obey Ward identities
\begin{equation}
    \qty[q^A{}_b(\omega,k)\mathcal{Y}^{bc}(\omega,k;\coes)]_{N+n+1}=\qty[S^{Ac}(\omega,k) P_\text{hy}(\omega,k;\coes)]_{N+n+1}.
\end{equation}

\section{Group structure}
\label{app:algebra}
As noted, the action of frame and on-shell transformations in the space of transport coefficients has the same form (see Eqs.~\eqref{app:coesF-action} and \eqref{app:coesO-action}), namely,
\begin{equation}    \coes'=f_{\Xnone} (\coes) \Rightarrow c_{\fatj}'(\omega,k)=\qty[c_{\fatj}+\sum_{\fati}x_{\fati}\qty((\gamma_{\fati})_{\fatj}+\sum_{\fatk}(\sigma_{\fati})_{\fatj\fatk}c_{\fatj})]_{n-n_\fatj}.
\end{equation}
The vectors $\gamma_\fati(\omega,k)$ and matrices $\sigma_\fati(\omega,k)$ obey (see Eqs.~\eqref{app:sigmaleftalgebra} and \eqref{app:sigmarightalgebra})
\begin{subequations}
    \begin{align}
        \sigma_\fati \sigma_\fatj = \sum_\fatk \rho_{\fati\fatj\fatk}\sigma_\fatk,\\
        \sigma_\fati \gamma_\fatj = \sum_\fatk \rho_{\fati\fatj\fatk}\gamma_{\fatk},
    \end{align}
\end{subequations}
for a set of coefficients $\rho_{\fati\fatj\fatk}(\omega,k)$. By introducing matrices $(\rho_\fati)_{\fatj\fatk}=\rho_{\fatj\fati\fatk}$, we also get (see Eqs.~\eqref{app:rholeftalgebra} and \eqref{app:rhorightalgebra})
\begin{equation}
    \rho_\fati \rho_\fatj =\sum_\fatk \rho_{\fati\fatj\fatk}\rho_\fatk.
\end{equation}
These transformation obey the composition rule
\begin{equation}
    f_{\Xnone'} \circ f_{\Xnone} = f_{\Xnone''} \Rightarrow x''_\fatk=\qty[x'_\fatk+x_\fatk+x'_\fati x_\fatj \rho_{\fati\fatj\fatk}]_{n-q_\fatk}.
\end{equation}
Here $q_\fatj$ is either $\vec{q}_\fatj$ or $\cev{q}_\fatj$, depending on whether we are working with on-shell or frame transformations. To express the inverse, we first introduce a matrix $R(\Xnone)$ as
\begin{equation}
    R_{\fati\fatj}(\Xnone)=\delta_{\fati\fatj}+\sum_\fatk x_\fatk \rho_{\fatj\fatk\fati}.
\end{equation}
The inverse transformation $\Xnone'$ is given by
\begin{equation}
    f_{\Xnone'}=(f_{\Xnone})^{-1} \Rightarrow x'_\fati = -\qty[(R^{-1}(\Xnone))_{\fati\fatj} x_\fatj]_{n-q_\fati}.
\end{equation}

We define the group $H$ through the action of frame and on-shell transformations on the space of transport coefficients, where we must identify all the transformations which may result in the same action. From Eqs.~\eqref{app:fr:Jaction} and \eqref{app:OSdef} it can be concluded that the frame and on-shell transformations commute, which gives
\begin{equation}
    \sigmaright_\fati \sigmaleft_\fatj = \sigmaleft_\fatj \sigmaright_\fati. 
\end{equation}
The properties \eqref{algebra:commutator} and \eqref{algebra:normal} then follow trivially.

We define the subgroup of all $m$-th order transformations $H_m$ as the set of all transformations for which
\begin{equation}
    f \in H_m: \qquad \qty[\CJ^a(f(\coes))]_{m-1}=\qty[\CJ^a(\coes)]_{m-1}.
\end{equation}
Our goal is to show that the commutator of an $m_1$-th order and an $m_2$-th order transformation is an $(m_1+m_2)$-th order transformation. Specifically, we want to show that for $f \in H_{m_1}$ and $g \in H_{m_2}$,
\begin{equation}
    f \circ g \circ f^{-1} \circ g^{-1} \in H_{m_1+m_2}
\end{equation}
Perhaps the easiest way to do this is to encode the transformations \eqref{app:fr:Jaction} and \eqref{app:OSdef} into block matrices $\vec{A}$ and $\cev{A}$ as
\begin{equation}
    \vec{A}(\Xright)=\mqty(\mathbb{I} + W(\Xright)q & -W(\Xright) \\ 0 & \mathbb{I}), \qquad \cev{A}(\Xleft)=\mqty(\Kleft(\Xleft) & \Omega(\Xleft) \\ 0 & \mathbb{I}),
\end{equation}
with $W(\Xright)$, $q$, $\Kleft(\Xleft)$ and $\Omega(\Xleft)$ as defined in Eqs.~\eqref{app:Wdef}, \eqref{app:qdef} and \eqref{app:KOmegadef}. These matrices faithfully encode all the information about the on-shell and frame transformations, and obey the transformation rules
\begin{subequations}
\label{app:groupA}
\begin{align}
    \vec f_{\Xright'} \circ \vec f_{\Xright}=\vec f_{\Xright''} &\Leftrightarrow \vec A(\Xright'')=\qty[\vec A(\Xright')\vec A(\Xright)]_n, \\  
    \vec f_{\Xright'} = (\vec f_{\Xright})^{-1} &\Leftrightarrow \vec A(\Xright')=\qty[\vec A^{-1}(\Xright)]_n,\\
    \cev f_{\Xleft'} \circ \cev f_{\Xleft}=\cev f_{\Xleft''} &\Leftrightarrow \cev A(\Xleft'')=\qty[\cev A(\Xleft)\cev A(\Xleft')]_n, \\  
    \cev f_{\Xleft'} = (\cev f_{\Xleft})^{-1} &\Leftrightarrow \cev A(\Xleft')=\qty[\cev A^{-1}(\Xleft)]_n.
\end{align}    
\end{subequations}
Note that the composition rule is with respect to left multiplication for on-shell transformations, and with respect to right multiplication for frame transformations. Note also that Eqs.~\eqref{app:groupA} imply Eqs.~\eqref{eq:matrixprod} and \eqref{eq:matrixinverse}. Since frame and on-shell transformations commute, it is sufficient to consider them separately. On the level of the matrices introduced above, the $m$-th order on-shell transformations are simple to write down as
\begin{subequations}
\label{app:order-props}
    \begin{align}
        \vec f_{\Xright} \in H_m \Leftrightarrow \qty[\vec A(\Xright)]_{m-1}=\mathbb{I},\\
        \cev f_{\Xleft} \in H_m \Leftrightarrow \qty[\cev A(\Xleft)]_{m-1}=\mathbb{I}.
    \end{align}
\end{subequations}
The properties \eqref{eq:algebra-props} follow trivially from Eqs.~\eqref{app:order-props}. To continue, it is useful to consider for a moment two matrices, $A=\mathbb{I}+\lambda^{m_1}\alpha$ and $B=\mathbb{I}+\lambda^{m_2}\beta$, with inverses $A^{-1}=\mathbb{I}-\lambda^{m_1}\bar\alpha$ and $B^{-1}=\mathbb{I}-\lambda^{m_2}\bar\beta$, where $\lambda$ is an expansion parameter. It is simple to show that, for $m_1\geq 0$, $m_2\geq 0$
\begin{equation}
    ABA^{-1}B^{-1}=\mathbb{I}+\lambda^{m_1+m_2} (\alpha \bar \beta-\beta \bar \alpha)+\CO(\lambda^{m_1+m_2+1}) .\label{app:comm}
\end{equation}
Let us translate this to on-shell transformations. We can write
\begin{subequations}
\begin{align}
    \vec f_{\Xright} \in H_{m_1} &\Rightarrow \vec A(\Xright) = \mathbb{I}+\vec{\alpha}(\Xright), \quad\qty[\vec{\alpha}(\Xright)]_{m_1-1}=0,\\
    \vec f_{\Xright'} \in H_{m_2} &\Rightarrow \vec A(\Xright') = \mathbb{I}+\vec{\alpha}(\Xright'), \quad\qty[\vec{\alpha}(\Xright')]_{m_2-1}=0.
\end{align}    
\end{subequations}
The commutator of these transformations is given by
\begin{equation}
    \vec f_{\Xright''}=\vec f_{\Xright} \circ \vec f_{\Xright'} \circ (\vec f_{\Xright})^{-1} \circ (\vec f_{\Xright'})^{-1} \Rightarrow \vec A(\Xright'')=\qty[\vec A(\Xright) \vec A(\Xright;)\vec A^{-1}(\Xright)\vec A^{-1}(\Xright')]_n. \label{app:xprimeprime}
\end{equation}
By taking $\omega\rightarrow\lambda\omega$ and $k\rightarrow\lambda k$, we can use Eq.~\eqref{app:comm} to conclude that
\begin{equation}
    \qty[\vec A(\Xright'')]_{m_1+m_2-1}=\mathbb{I},
\end{equation}
which, in turn, means that $\vec f_{\Xright''}$ as defined in Eq.~\eqref{app:xprimeprime} is a transformation of $(m_1+m_2)$-th order. A completely analogous computation can be performed also for the frame transformations, using matrices $\cev A(\Xleft)$. The rest of the properties of $H$, described in Section~\ref{sec:algebra}, then follow.

\twocolumngrid
\bibliography{biblio}

\end{document}